\documentclass[aps,twocolumn,prd,superscriptaddress,nofootinbib]{revtex4-1}

\usepackage{mathtools}
\usepackage{amsfonts}
\usepackage{mathrsfs}
\usepackage{bbm}
\usepackage{slashed}

\usepackage{graphicx}
\usepackage{color}
\usepackage{array}

\usepackage{placeins}
\usepackage{booktabs}
\usepackage[caption=false]{subfig}

\usepackage{xspace}
\usepackage{siunitx}
\usepackage{hyperref}
\usepackage[nameinlink]{cleveref}
\usepackage{bookmark}

\usepackage{xifthen}
\usepackage{xcolor}
\hypersetup{
	colorlinks,
	linkcolor={red!75!black},
	citecolor={blue!75!black},
	urlcolor={blue!75!black}
}
\usepackage{units}

\usepackage[utf8]{inputenc}

\setkeys{Gin}{width=0.48\textwidth}
\newcommand*{\ie}{i.e.\@\xspace}

\newcommand*{\cf}{cf.\@\xspace}

\newcommand{\imag}{\text{i}}

\newcommand{\gettitle}{Spectral functions in the $\phi^4$-theory from the spectral DSE}

\newcommand{\getHeidelbergAffiliation}{\affiliation{Institut f\"ur Theoretische Physik, Universit\"at Heidelberg, Philosophenweg 16, 69120 Heidelberg, Germany}}

\newcommand{\getEMMIAffiliation}{\affiliation{ExtreMe Matter Institute EMMI, GSI, Planckstr. 1, 64291 Darmstadt, Germany}}

\hypersetup{
	colorlinks,
	linkcolor={red!75!black},
	citecolor={blue!75!black},
	urlcolor={blue!75!black},
	pdftitle={\gettitle},
	pdfauthor={Horak,Pawlowski, Wink},
	pdfkeywords={analytic continuation}
	{correlations functions} {scalar theory}
	{functional renormalisation group}
	{real time} {spectral function} 
	bookmarksopen=true,
	bookmarksopenlevel=2,
	bookmarksnumbered=true
}

\begin{document}

\title{\gettitle}

\author{Jan Horak}
\getHeidelbergAffiliation

\author{Jan M. Pawlowski}
\getHeidelbergAffiliation
\getEMMIAffiliation
  
\author{Nicolas Wink}
\getHeidelbergAffiliation

\begin{abstract}
We develop a non-perturbative functional framework for computing real-time correlation functions in strongly correlated systems. The framework is based on the spectral representation of correlation functions and dimensional regularisation. Therefore, the non-perturbative spectral renormalisation setup here respects all symmetries of the theories at hand. In particular this includes space-time symmetries as well as internal symmetries such as chiral symmetry, and gauge symmetries. Spectral renormalisation can be applied within general functional approaches such as the functional renormalisation group, Dyson-Schwinger equations, and two- or $n$-particle irreducible approaches. As an application we compute the full, non-perturbative, spectral function of the scalar field in the $\phi^4$-theory in $2+1$ dimensions from spectral Dyson-Schwinger equations. We also compute the $s$-channel spectral function of the full $\phi^4$-vertex in this theory.       
\end{abstract}

\maketitle

\section{Introduction} \label{sec:Introduction}

The study of the dynamics and resonance structure of strongly correlated systems requires the knowledge of real-time (time-like) correlation functions. In particular the evolution of slow modes is dominated by the low-energy regime and is crucial for transport approaches and hydrodynamics. Spectral functions encode the full, non-perturbative, information of the respective degrees of freedom and open the door to additional real-time quantities such as transport coefficients. They are also a particularly useful tool when discussing resonances and bound states, since they give direct access to the spectrum of excitations in a given theory. Emergent composite states are of central interest not just in particle and nuclear physics, but in most physics areas. 

The treatment of strongly correlated systems asks for non-perturbative techniques. In recent years functional and lattice approaches have been applied very successfully to the low-energy regime of theories ranging from QCD to condensed matter systems. The general framework for the quantum-field theoretical description of bound states with functional methods was introduced by Bethe and Salpeter~\cite{Salpeter:1951zz, Salpeter:1951sz}. For a recent review see \cite{Eichmann:2016yit}, for a review of respective lattice results see~\cite{Aoki:2016frl}.

Despite this rapid and very impressive progress in particular in the last decade, the reliable direct non-perturbative calculation of real-time correlation functions is still in its infancy. Functional approaches have matured  in recent years and are by now quantitatively competitive in QCD in Euclidean space-time. The extension of such calculations to Minkowski space-time is yet hindered by technical and conceptual complications, for recent advances see~\cite{Krein:1993jb, Floerchinger:2011sc, Strauss:2012dg, Dorkin:2013rsa, Tripolt:2013jra, Dorkin:2014lxa, Pawlowski:2015mia, Dorkin:2015jck, Yokota:2016tip, Kamikado:2016chk, Jung:2016yxl, Pawlowski:2017gxj, Jia:2017niz, Yokota:2017uzu, Wang:2017vis, Bluhm:2018qkf, Tripolt:2018jre, Tripolt:2018qvi, Wang:2018osm, Alkofer:2018guy, Eichmann:2019dts, Solis:2019fzm, Aarts:2001qa, Ivanov:2005bv, roder2006self, Marko:2015gpa, Shen:2019jhl, Shen:2020jya}.

In this work we develop a novel approach for the direct computation of real-time (Minkowski) observables  that is based on spectral representations of correlation functions. The approach comes with the advantage that we can use dimensional regularisation for the analytic computation of momentum integrals in fully numerical non-perturbative calculations. Accordingly, the respective renormalisation scheme, \textit{spectral renormalisation}, is based on  standard dimensional regularisation and respects the space-time symmetries, internal symmetries such as chiral symmetry, and gauge symmetries of the theory at hand. Clearly, this method is applicable to a broad range of theories, including non-Abelian gauge theories, within a regularisation and renormalisation scheme which is manifestly gauge-invariant. 

In the present work we apply our novel approach to the scalar $\phi^4$-theory in $d=2+1$ dimensions. This theory is a simple strongly correlated system and serves as a good benchmark for new techniques before applying them more involved theories such as non-Abelian ones. It is also interesting in its own right and has many applications as a model theory for perturbative and non-perturbative phenomena. 

The numerical application in the present work is done within the spectral Dyson-Schwinger approach. We compute the spectral function of the scalar field from the gap equation for the propagator, using its K{\"a}llen-Lehmann spectral representation. In a first step all vertices are approximated with the classical ones, and the two-loop diagrams in the Dyson-Schwinger equations (DSE) are included. In a second step a use a skeleton expansion of the DSE with a bubble-resummation of the $s$-channel  four-point function. The non-perturbative $s$-channel spectral function of the vertex is computed and used in the DSE. The setup gives us direct numerical access to Minkowski space-time correlation functions due to the spectral representation. This also allow for a (physical)  on-shell renormalisation via the \textit{spectral renormalisation} scheme. 

The paper is organized as follows: In~\Cref{sec:SpecRen} we introduce \textit{spectral renormalisation} at the example of the  Dyson-Schwinger approach along with the necessary technical tools. In particular we discuss \textrm{dimensional} \textrm{spectral} \textrm{renormalisation} and a  Bogoliubov-Parasiuk-Hepp-Zimmermann--type (BPHZ) spectral renormalisation. \Cref{sec:Calculation} works out the calculational details of solving the DSE in the developed scheme, followed by our results in~\Cref{subsec:results_bare_vertices} for classical vertices and in~\Cref{subsec:res4ptfct}-\Cref{subsec:self_consistent_polarisation} for the skeleton expansion. Finally, our conclusion is presented in \Cref{sec:Conclusion}.

\section{Spectral renormalisation}
\label{sec:SpecRen}
\begin{figure}[t]
	\centering
	\subfloat{\includegraphics[width=0.15\textwidth]{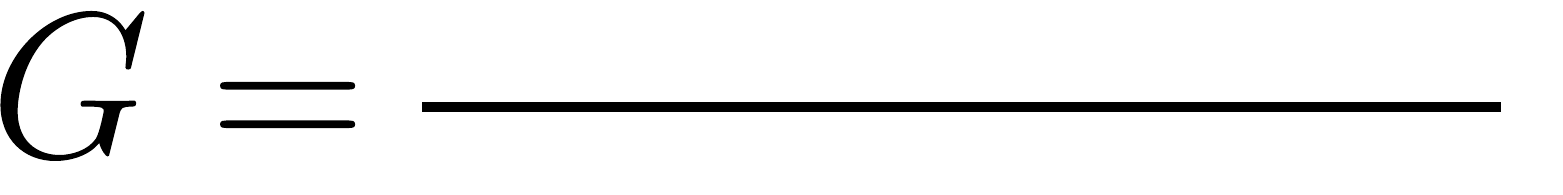}}\\
	\subfloat{\includegraphics[width=0.15\textwidth]{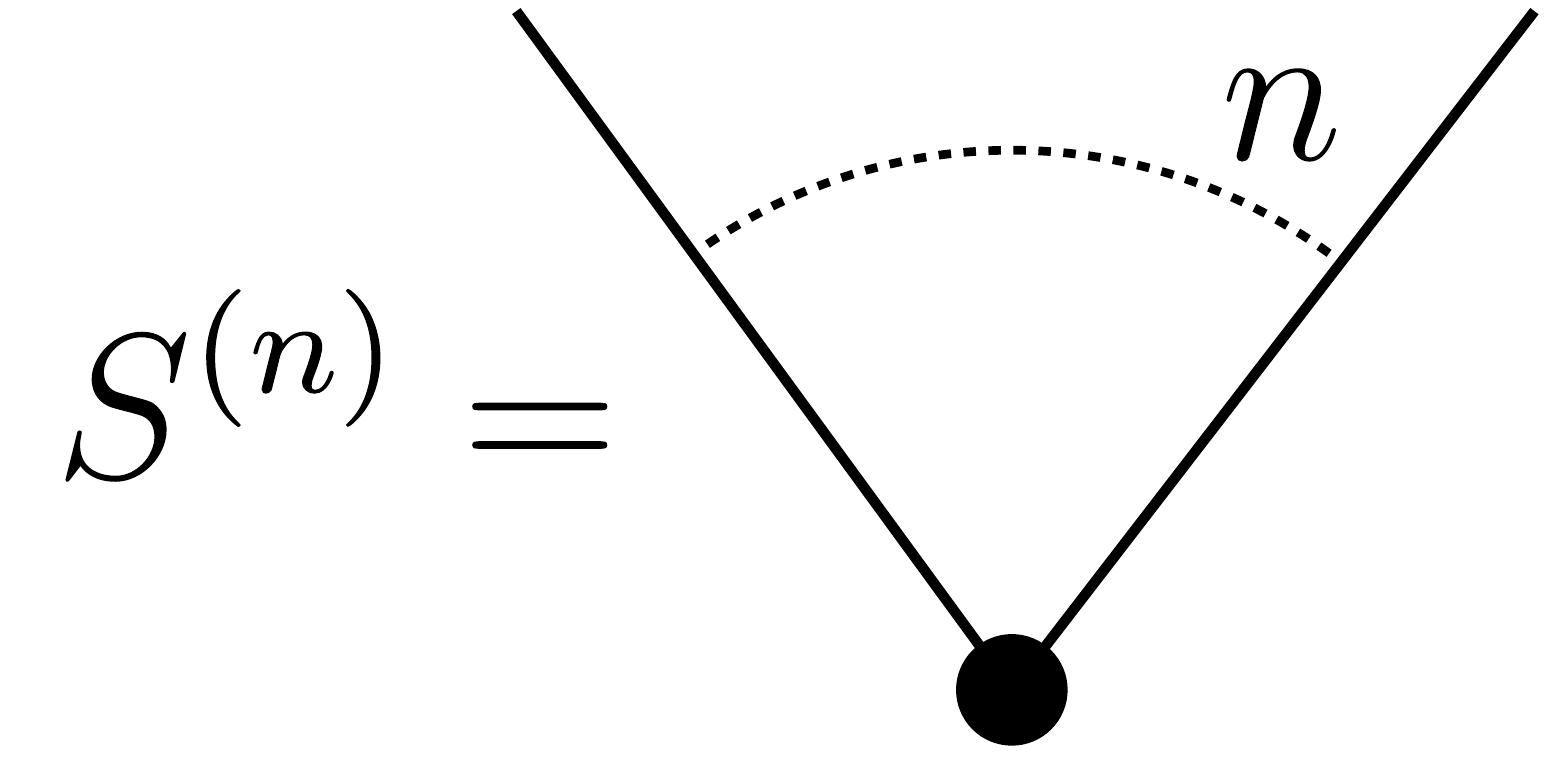}}\hspace{1cm}
	\subfloat{\includegraphics[width=0.15\textwidth]{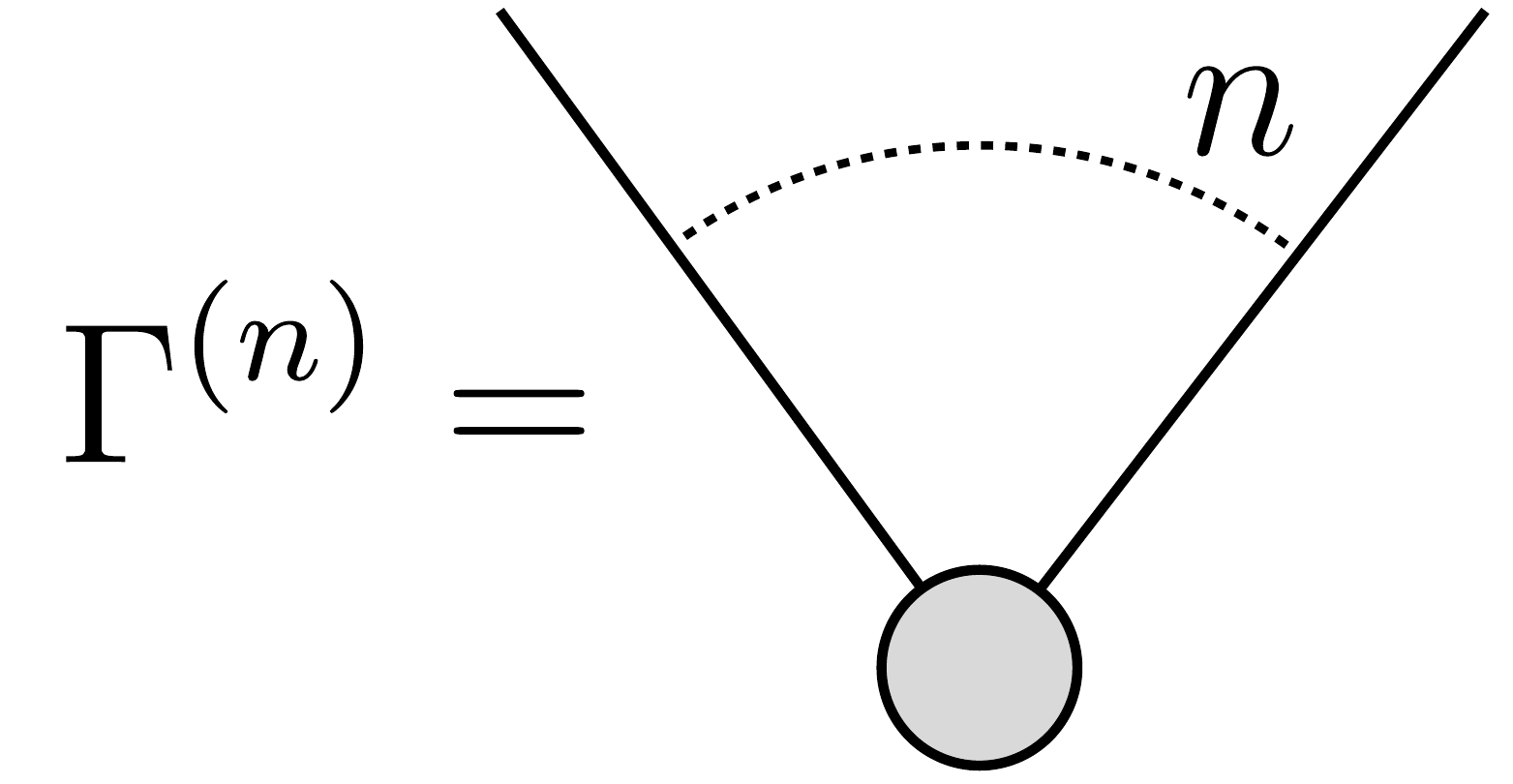}}
	\caption{Diagrammatic notation used throughout this work: Lines stand for full propagators, small black dots stand for classical vertices, and larger grey dots stand for full vertices. }
	\label{fig:FunctionalMethods:DSE_notation}
\end{figure}
The general real-time renormalisation scheme developed here aims at combining a practical numerical implementation in non-perturbative applications while maintaining all underlying symmetries including gauge symmetries. This is achieved by utilising dimensional regularisation, which respects all space-time, internal and gauge symmetries of the theory at hand. We also develop a BPHZ-type subtraction scheme which facilitates the analytical computations significantly. If such a subtraction schemes does not violate any symmetries in the theory at hand, it is the scheme of choice. 

A practical implementation of dimensional regularisation requires an analytic momentum structure of the propagators and vertices in the given loop integrals. While this allows for its use in perturbation theory, non-perturbative applications, with their necessarily numerical computation of propagators and vertices, usually rely on ultraviolet momentum cutoffs. The latter are neither consistent with space-time symmetries nor with gauge symmetries. It is well-known that in gauge theories and supersymmetric theories such a regularisation requires symmetry-breaking  counterterms. This is not a conceptual problem, but it typically triggers additional power-counting relevant terms that may leads to additional fine-tuning tasks, see e.g.~\cite{Cyrol:2016tym, Cyrol:2017ewj}. Moreover, the Wick rotation to Minkowski space-times is hampered by the deformation of the momentum integrals, which can lead to additional poles and cuts in the integration contours. 

The present renormalisation scheme achieves the requirement of analytic momentum integrals by using spectral representations of correlation functions. For the propagators this is the K{\"a}llen-Lehmann spectral representation, similar spectral representations also exist for the vertices, though getting increasingly difficult. We call this renormalisation scheme  \textit{spectral renormalisation}: after inserting the spectral representations in the loop integrals, the momentum integrands take an analytic form. This form is well-suited for using dimensional regularisation or related analytic computation techniques. The loop-momentum integrations can be performed analytically and we are left with spectral integrals. The whole non-perturbative information is contained in the spectral functions of propagators and vertices. In most non-perturbative applications the respective spectral integrals can only be computed numerically.

\begin{figure}[t]
	\centering
	\includegraphics[width=0.85\linewidth]{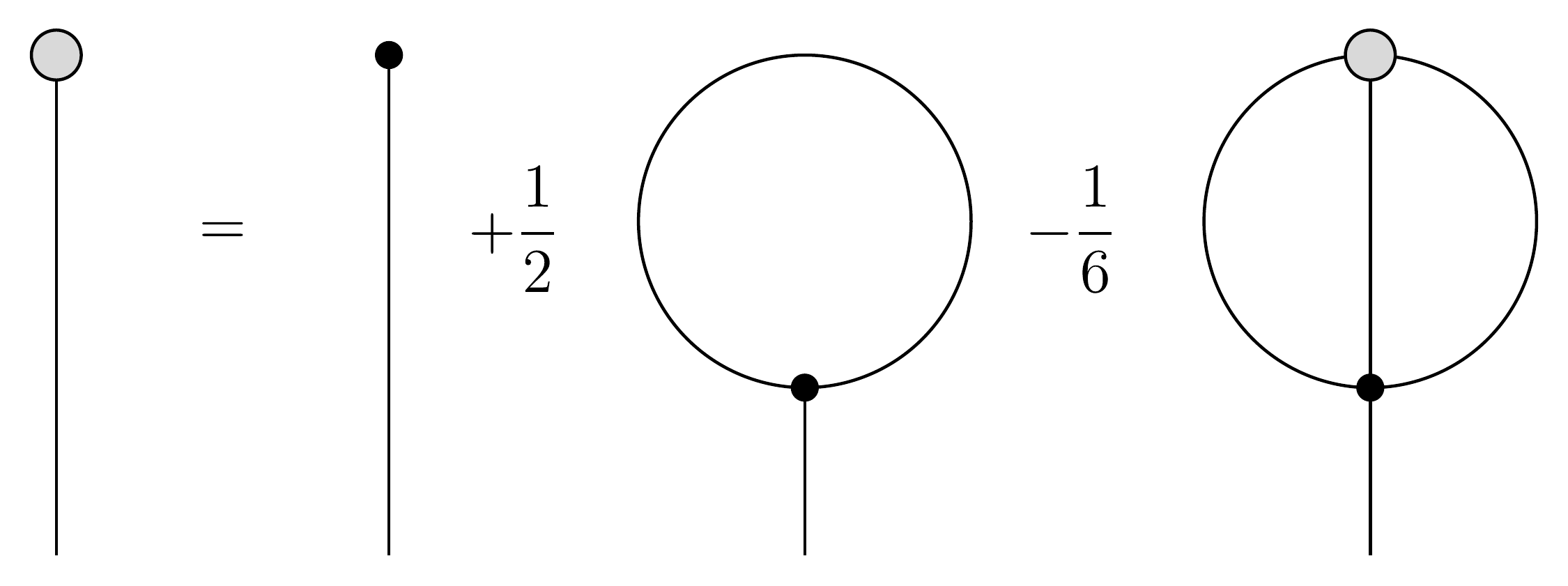}
	\caption{Functional DSE for the effective action of the scalar theory under investigation in this work. The notation is given in \Cref{fig:FunctionalMethods:DSE_notation}.}
	\label{fig:DSE_1pt}
\end{figure}
\begin{figure*}[t]
	\centering
	\includegraphics[width=.95\linewidth]{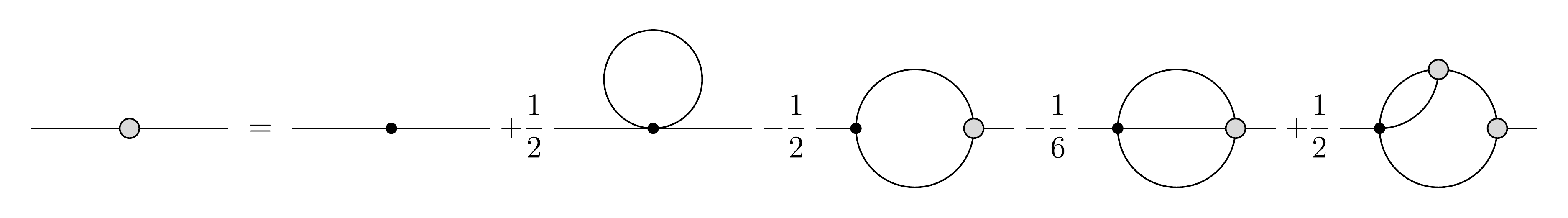}
	\caption{DSE of the two-point function for a general background field $\phi\neq 0$. The vacuum polarisation and squint diagrams are proportional to $S^{(3)}[\phi]\propto \phi$. They vanish for $\phi=0$, where the standard form with tadpole and sunset diagrams is obtained. The notation is given in \Cref{fig:FunctionalMethods:DSE_notation}.  }
	\label{fig:DSE_2pt}
\end{figure*}

The scheme can be practically applied to any divergent diagram that scales in the UV with loop momentum to some natural power $q^m$, $m \in \mathbb{N}$ (with $m < n_{\text{max}}$ and $n_{\text{max}}$ given by the renormalisability constraint). This is always the case when using spectral representations for all correlation functions, but also works for  classical vertices. This will be detailed in the present work within the example of the Dyson-Schwinger approach introduced in the next section, \Cref{subsec:dse}. The functional DSE for  the effective action,  \labelcref{eq:DSE}, is depicted in \Cref{fig:DSE_1pt}, that for the inverse propagator is depicted in \Cref{fig:DSE_2pt}. For classical vertices, all momentum integrals in functional approaches, see e.g.~\Cref{fig:DSE_2pt}, are of the standard perturbative form, but with different spectral masses for all lines.  Most of these integrals are known from perturbation theory results, e.g.~\cite{Rajantie:1996np}. Note that this re-parameterisation comes at the cost of spectral integrals for each propagator. Spectral representations of vertices lead to further spectral integrals as well as further classical propagators with spectral masses. In summary, in a spectral functional approach all momentum integrals are perturbative. Hence we can implement a symmetry-preserving regularisation such as dimensional regularisation, leading to a renormalisation scheme with symmetry-consistent counterterms. 

A relevant example for this important symmetry-preserving property are gauge theories. There, a momentum cutoff or standard subtraction scheme requires explicitly or implicitly a mass counterterm for the gauge field in order to keep the renormalised gauge field massless. In four dimensions this leads to a quadratic fine-tuning task instead of a logarithmic one, for a detailed discussion see \cite{Cyrol:2016tym, Cyrol:2017ewj}. \textit{Spectral renormalisation} with spectral regularisation removes the (explicit or implicit) necessity of a mass counterterm for the gauge field, and hence the quadratic fine-tuning task. 

After briefly introducing Dyson-Schwinger equations, \Cref{subsec:dse}, and the K\"allen-Lehman spectral representation, \Cref{subsec:spec_rep}, we set up \textit{spectral renormalisation} in \Cref{subsec:spectral_renormalisation} (spectral dimensional renormalisation), and \Cref{subsec:SpecBPHZ} (spectral BPHZ-renormalisation). Further examples and the discussion of the fully non-perturbative setup can be found in \Cref{subsubsec:spectral_renormalisation_one_loop} and \Cref{subsec:spectral_renormalisation_beyond_one_loop}. The explicit example used for demonstrating the properties and computational details of the spectral renormalisation scheme is the gap equation for scalar $\phi^4$- and $\phi^3$-theories. 

\subsection{Dyson-Schwinger equations}
\label{subsec:dse}


The central object in functional approaches such as Dyson-Schwinger equations or functional renormalisation group equations is the quantum effective action $\Gamma[\phi]$. It is related to the generating functional $Z[J]$ of correlation functions including their disconnected parts via a Legendre transformation, 
\begin{equation}
\label{eq:gamma}
	\Gamma[\phi] = \sup_J\left[ \int d^dx \,J(x)\phi(x) - W[J]\right]
\, .
\end{equation}
Here, $W[J] = \ln Z[J]$ is the Schwinger functional that generated connected correlation functions. The effective action $\Gamma[\phi]$ is the generating functional for one-particle irreducible (1PI) correlation functions. They are obtained from $n$th derivative of the effective action w.r.t.\ to the fields. In momentum space this reads,   
\begin{equation}
\label{eq:gamma_n}
	\Gamma^{(n)}[\phi](p_1,\dots,p_n) = \frac{\delta^n\Gamma[\phi]}{\delta\phi(p_1)\dots\delta\phi(p_n)}
\, .
\end{equation} 
A pivotal role in all functional approaches is played by the full, field-dependent propagator $G[\phi]$. It is simply the inverse of the 1PI two-point function, 
\begin{align}
\label{eq:prop_gamma_fourier}
	G[\phi](p,q) = \frac{1}{\Gamma^{(2)}}[\phi](p,q)
\, .  
\end{align} 
With these prerequisites we straightforwardly arrive at the master Dyson-Schwinger equation (DSE), the quantum equation of motion, 
\begin{align}
\label{eq:DSE}
	\frac{\delta\Gamma[\phi]}{\delta\phi} = \frac{\delta S}{\delta\varphi}\left[\varphi=G\cdot\frac{\delta}{\delta\phi}+\phi\right]
\, ,
\end{align}
where $G\cdot\delta/\delta\phi = \int d^d q/(2 \pi)^d G(p,q) \delta/\delta\phi(q)$ in momentum space, and $S[\phi]$ is the classical action. A more detailed derivation can be found e.g.\ in~\cite{Alkofer:2000wg}. By taking functional derivatives of~\labelcref{eq:DSE} with respect to $\phi$, all higher order 1PI correlation functions are generated. For example, the DSE for the two-point function is generated by taking one derivative of \labelcref{eq:DSE}, and is depicted in \Cref{fig:DSE_2pt}. 

For our explicit example of a $\phi^4$-theory, the classical action in \labelcref{eq:DSE} is given by, 
\begin{align}
\label{eq:scalar_action}
	S[\varphi]=\int d^dx \left[ \frac{1}{2}(\partial_\mu\varphi)^2 + \frac{m_{\phi,0}^2}{2} \varphi^2 + \frac{\lambda_{\phi,0}}{4!} \varphi^4 \right]
\, . 
\end{align}
It depends on two parameters or couplings, the bare four-point coupling $\lambda_{\phi,0}$ and the bare mass parameter $m_{\phi,0}$. The third parameter required for renormalisation, the wave function renormalisation $Z_\phi$ can be scaled out. This is conveniently done by setting it to unity at the renormalisation scale. 

We close this section with a few comments: To begin with,  the underlying $Z_2$-symmetry of the theory, $\varphi \rightarrow - \varphi$ implies the same symmetry for the effective action under transformations of the mean field, $\phi \rightarrow - \phi$. Accordingly, the odd vertices vanish at vanishing mean field: $\Gamma^{(2n+1)}[\phi=0] \equiv  0$. Moreover, restricting ourselves to constant background fields $\phi_c$, we can formally expand the three-point function in powers of the field, 
\begin{equation}
\label{eq:three_point_func}
	\Gamma^{(3)}[\phi_c](p_1,p_2,p_3) = \phi_c \bigg[ \Gamma^{(4)}(p_1,p_2,p_3,0) + O(\phi_c^2)\bigg] \,, 
\end{equation}
due to the odd vertices vanishing at the expansion point $\phi = 0$. In \labelcref{eq:three_point_func} we have used the Fourier transform $\tilde \phi_c(p) = (2 \pi)^d \phi_c\delta(p)$ of the constant field. We have also introduced $\Gamma^{(n)}(p_1,p_2,p_3,0,...,0)$, the $n$-point functions at a vanishing background and $n-3$ vanishing momenta. The relation \labelcref{eq:three_point_func} gives rise to the polarisation and squint diagram in \Cref{fig:DSE_2pt} for $\phi_c\neq 0$, which are absent for $\phi_c=0$. In the broken phase the equation of motion (EoM) $\phi_0$ for constant fields is solved for a non-vanishing expectation value of the field, i.e.~$\phi_0 \neq 0$. Accordingly, if evaluating the DSEs for correlation functions on the EoM, these diagrams are present. 

\subsection{K\"allen-Lehmann spectral representation}
\label{subsec:spec_rep}

Using the K{\"a}llen-Lehmann spectral representation~\cite{Kallen:1952zz,Lehmann:1954xi}, the propagator can be recast in terms of its spectral function $\rho$, 
\begin{equation} \label{eq:kaellenlehmann}
	G(p_0) = \int_0^\infty \frac{d\lambda}{\pi}\frac{\lambda\,\rho(\lambda,|\vec p|)}{p_0^2+\lambda^2} \,.
\end{equation}
For asymptotic states, the spectral function can be understood as a probability density for the transition to an excited state with energy $\lambda$. In this way, the spectral function acts as a linear response function of the two-point-correlator, encoding the energy spectrum of the theory. The existence of a spectral representation imposes tight restrictions on the analytic structure of the propagator. In turn, the Euclidean propagator also constrains the spectral function, for a rather non-trivial example for the latter constraints see \cite{Cyrol:2018xeq}.

From a complex analysis perspective, the spectral function naturally arises as the set of non-analyticities of the propagator, which are severely restricted in the complex plane by Cauchy's Theorem. This results into the following inverse relation between spectral function and the retarded propagator,
\begin{equation} \label{eq:spec_func_def}
\rho(\omega,\left|\vec{p}\right|) = 2 \; \text{Im} \; G(-\text{i} (\omega+\text{i} 0^+),\left|\vec{p}\right|) \,,
\end{equation} 
where $\omega$ is the real time zero momentum component. This formulation allows us to work only with the frequency argument and set the spatial momentum to zero in practice, since the full phase-space can be restored from Lorentz invariance. Hence, for the remainder of this work, $\left|\vec{p}\right|$ will be dropped.

The existence of a spectral representation restricts all non-analyticities of the propagator to lie on the real momentum axis, as is manifest in~\labelcref{eq:kaellenlehmann} and~\labelcref{eq:spec_func_def}. This crucial condition as well as the generic structure in the complex plane already mentioned above allow us to recast the spectral function into the form
\begin{equation} \label{eq:spec_func_expl}
\rho(\lambda) = \frac{\pi}{\lambda}\,\sum_i  Z_i\,\delta(\lambda-m_i)\;+\;\tilde{\rho}(\lambda) \,.
\end{equation}
The spectral function is split into a set of $\delta$-functions and a continuous scattering part $\tilde{\rho}$, arising from branch cuts in the complex plane of the propagator. For a classical propagator the spectral function reduces to one mass-shell $\delta$-function with $Z_1=1$. There are no further poles, $Z_{i>1}=0$, and the scattering part is absent, $\tilde{\rho}=0$. Further details can be found e.g.\  in~\cite{Peskin:1995ev}.

In the scalar $\phi^4$-theory, the spectral function of the scalar field is that of an asymptotic state, and hence is positive semi-definite and has the interpretation of a probability density. Therefore it is convenient to normalise its integrated weight to unity within an appropriate renormalisation scheme. Within this scheme we have the sum rule, 
\begin{equation} \label{eq:scalarsumrule}
\int_\lambda \lambda\; \rho(\lambda)=1 \quad\mathrm{with}\quad \int_{\lambda} \equiv \int_0^\infty \frac{d\lambda}{\pi}\,, 
\end{equation} 
which implies $Z_i\leq 1$. The spectral weight is distributed between poles and cuts, and in the presence of scattering states the weight of the poles is less than one. 

\subsection{Spectral dimensional renormalisation} \label{subsec:spectral_renormalisation}
\begin{figure}[t]
	\centering
	\includegraphics[width=\linewidth]{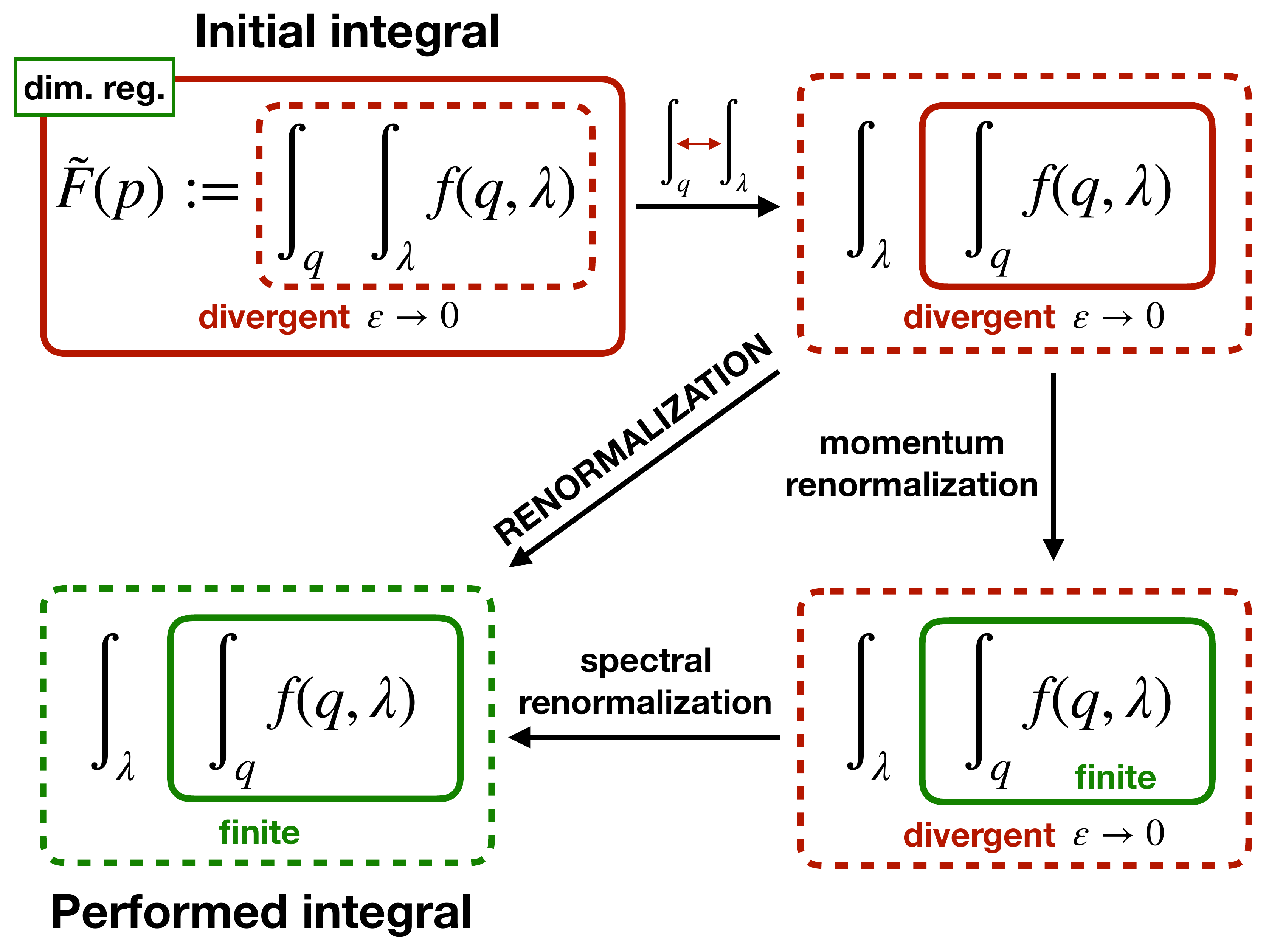}
	\caption{Schematic illustration of the spectral renormalisation scheme. The function $f$ is some arbitrary divergent integrand. The dependence on the external momentum $p$ is suppressed. The upper two boxes have to be understood as finite by dimensional regularisation, but divergent in the limit $\varepsilon \to 0$ in $d - \varepsilon$ dimensions. In a first step, the momentum integrals are analytically evaluated via dimensional regularisation. Subsequently, the spectral integrals are renormalised via spectral renormalisation, either within the dimensional-~(\Cref{subsec:spectral_renormalisation}) or BPHZ-approach~(\Cref{subsec:SpecBPHZ}).}
	\label{fig:integration_swap_scheme}
\end{figure}

Next, we discuss the \textit{spectral renormalisation} scheme, which is fully based on dimensional regularisation. 
Dimensional regularisation renders the loop diagrams finite and we can swap orders of the initially outer momentum and inner spectral integrals integrals (cf. upper part of~\Cref{fig:integration_swap_scheme}. This allows us to first perform the momentum integrals analytically. We are left with finite spectral integrals at finite $\epsilon > 0$, that in general have to be done numerically since the spectral functions may only be known numerically. 

Generally, the numerical integration of the spectral integrals can be performed for finite $\epsilon$, and in gauge theories full manifest gauge-consistency of \textit{spectral renormalisation} requires that the limit $\epsilon\to 0$ is taken only after performing all integrals. However, it is convenient for the numerical performance to do the spectral integration at $\epsilon=0$. The same holds for the access  to the analytic momentum structure required for extracting Minkowski properties. In this case, for the limit $\epsilon\to 0$ we have to set up a consistent renormalisation procedure before performing the spectral integrals, which is worked out in detail in~\Cref{subsec:SpecBPHZ}. In particular it is not sufficient to only remove the $1/\epsilon$-divergences that originate in the momentum integrations, as the spectral integrations lead to further $1/\epsilon$-terms. This relates to the swapping of the integration order and performing the limit  $\epsilon\to 0$ before evaluating all integrals. 

This section is dedicated to the fully gauge-consistent renormalisation scheme we call \textit{spectral dimensional renormalisation}. In this scheme, the divergent parts of the spectral integrals are performed analytically before taking the limit $\varepsilon \to 0$. A simple example is the tadpole contribution to the gap equation in \Cref{fig:DSE_2pt} in $d=3$ dimensions, which comes with a momentum-independent linearly divergent term. This example is also relevant for our later computation in \Cref{sec:Calculation} and \Cref{sec:results}. After the momentum integration is performed, we arrive at a finite result proportional to 
\begin{align}
\label{eq:SpectralSing} 
\int_0^\infty d\lambda\, \lambda  \frac{\mu^{2\epsilon} \lambda^{1-2\epsilon}}{(\lambda^2 +m^2)}= 
-\frac {\pi}{2}\frac{1}{\cos\left(\frac{\pi\epsilon}{2}\right)} \,m\left(\frac{\mu^2}{m^2}\right)^{\epsilon}\,. 
\end{align}
In \labelcref{eq:SpectralSing} we have used a trial spectral function that decays for large spectral values $\lambda$ according to its momentum or spectral dimension, 
\begin{align}
\label{eq:rhotrial}
\rho_\textrm{trial}(\lambda,m) = \frac{1}{\lambda^2+m^2}\,,
\end{align}
with a positive mass $m>0$. The trial spectral function in \labelcref{eq:rhotrial} approximates the correct leading ultraviolet behaviour, if we neglect logarithmic corrections. For large spectral values, the UV-asymptotic of the spectral function can be extracted from the leading momentum dependence of the respective propagator. In the present example, the latter is assumed to decay quadratically on its branch cut on the real momentum axis. 

The finiteness of the result of the momentum integration used on the left hand side of \labelcref{eq:SpectralSing}  also points at a specific property of dimensional regularisation: in odd dimensions,  $d=2 n +1$, it already removes all  momentum divergences. In even dimensions, $d=2n$, it removes subclasses of divergent ones, a prominent being the (one-loop) tadpole in a massless theory such as a gauge theory.
 
If we had put $\epsilon=0$ before the integration, \labelcref{eq:SpectralSing} simply is (linearly) divergent. This is the price to pay for the swapping of integration orders: the divergences are not fully covered by the momentum integrals any more. The example also entails that in odd dimensions including our explicit computation in the $\phi^4$-theory in $d=3$, \textit{all} divergences come via the spectral integrals. 

In \textit{spectral dimensional renormalisation}, spectral singularities in even dimensions show up as $1/\epsilon$-terms. In dimensional regularisation in perturbation theory these divergences are typically removed recursively by introduction of appropriate counterterms. In spectral dimensional renormalisation this procedure is applied too, while keeping a finite $\epsilon$ as well as isolating analytically the singular part of the spectral integrals with 
\begin{align}\label{eq:IR-UV-split}
\rho(\lambda)= \rho_{\textrm{IR}}(\lambda)+ \rho_{\textrm{UV},\textrm{an}}(\lambda)\,.
\end{align}
In \labelcref{eq:IR-UV-split}, the numerical 'infrared' part $\rho_{\textrm{IR}}(\lambda)$ decays sufficiently fast for large spectral values and renders the respective spectral integrals finite. In turn, the ultraviolet part $\rho_{\textrm{UV},\textrm{an}}(\lambda)$ carries the ultraviolet asymptotics analytically. Therefore, the respective spectral integrals can be treated analytically with dimensional regularisation as done in \labelcref{eq:SpectralSing}. With \labelcref{eq:rhotrial} we use the IR-UV--split 
\begin{align}\label{eq:IR-UVsplit}
\rho(\lambda) = \rho_\textrm{IR}(\lambda,k )+ \rho_\textrm{trial}(\lambda,k)\,. 
\end{align}				
We also emphasise that the theory does not depend on the mass parameter $k$ that regularises (in the infrared) the UV-part of the spectral function. In particular we have $\partial_k\rho(\lambda)\equiv 0$, which entails the $k$-dependence of the infrared part of the spectral function.

The (leading) ultraviolet behaviour of the spectral function at large spectral values is governed by $\rho_{\textrm{UV},\textrm{an}} = \rho_{\textrm{trial}}$ and the infrared part decays with the fourth power of the spectral value, 
\begin{align}
	\label{eq:UVdecayrhoIR}	
	\lim_{\lambda \to\infty}\rho_\textrm{IR}(\lambda \to\infty)\propto \frac{1}{\lambda^4} 
\end{align}
Inserting the split \labelcref{eq:IR-UVsplit} into \labelcref{eq:SpectralSing}  leads us to the final finite result with $\epsilon=0$, 
\begin{align}
 \int_0^\infty d\lambda\,\mu^{2\epsilon} \lambda^{2-\epsilon} \rho(\lambda)=&\, 
 \int_0^\infty d\lambda\,   \lambda^{2} \rho_\textrm{IR}(\lambda) 
 -\frac {\pi}{2}  \,k\,, 
 \label{eq:SpecRenPractical}\end{align}
with a finite (in general numerical) integral over the infrared part of the spectral function due to \labelcref{eq:UVdecayrhoIR}. The numerical convergence of this integral can be further improved systematically, if the UV-part $\rho_\textrm{UV}(\lambda)$ also includes sub-leading UV-terms of the full spectral function. 

The finite result of \labelcref{eq:SpecRenPractical} was obtained without the introduction of any possibly symmetry-breaking counterterms. Since the systematics of this example is general, it can be applied to all divergences of general diagrams. As the demonstrated \textit{spectral dimensional renormalisation} procedure is entirely based on dimensional regularisation and the use of spectral representation, it preserves all symmetries of the theory at hand. Especially for the case of gauge theories it is manifestly gauge-invariant or rather gauge-consistent. For example, it reflects the peculiarity of dimensional regularisation that integrals without any external scale vanish identically. For $k=0$ the integral over the UV-part of the spectral function vanishes and we are left with the finite IR-part. Accordingly, no mass counterterms are needed in a massless theory such as a gauge theory. 

\Cref{fig:integration_swap_scheme} illustrates the general renormalisation workflow, including spectral renormalisation. The schematic representation holds for the case of spectral dimensional renormalisation as well as for the subtraction-based and more universal approach of \textit{spectral BPHZ-renormalisation}, which is introduced in the following.

\subsection{Spectral BPHZ-renormalisation}\label{subsec:SpecBPHZ}

In fully symmetry-consistent \textit{spectral dimensional renormalisation} as described in \Cref{subsec:spectral_renormalisation}, one has to perform analytic spectral integrals with dimensional regularisation on top of the momentum integrations. Already the latter are more complicated than in standard perturbation theory at the same order, since the spectral representations lead to different masses for each line. While the momentum integration has to be necessarily analytic in order to access Euclidean and Minkowski space-time, this is not necessary for the spectral integration, whose non-perturbative infrared part has to be done numerically in most cases anyway.

The need for additional analytic computations of spectral integrals can be circumvented by performing subtractions on the spectral integrals which render the spectral integrals finite. This can be done by subtracting a Taylor expansion of the spectral integrand in momenta according to the BPHZ-scheme with Dyson's formula. The procedure is called \textit{spectral BPHZ-renormalisation}. We shall see, that its workflow is still described within \Cref{fig:integration_swap_scheme}. It is the last "spectral" step from the bottom right to the bottom left which is changed by moving from spectral dimensional to the spectral BPHZ-procedure. We emphasise that the underlying BPHZ-regularisation in general does not preserve all symmetries of a given theory and in particular breaks gauge symmetry. This is not a conceptual problem, as the counterterms also break gauge invariance and the final result is gauge-consistent, see e.g.~\cite{Kraus:1997bi}. However, in numerical applications to gauge theories the gauge-consistent \textit{spectral dimensional renormalisation} is arguably worth its price, in particular for investigations of the Gribov problem and the confinement mechanism.

In the present example of a scalar $\phi^4$-theory, the BPHZ-scheme is consistent with both space-time and the internal $Z_2$-symmetry. We will therefore utilise it for explicit computation. We introduce and explain the setup within a specific example, the sunset graph in the gap equation of the scalar $\phi^4$-theory in $d$ dimensions, see~\Cref{fig:DSE_2pt}. This diagram also carries sub-divergences while still being relatively simple. In \Cref{subsubsec:spectral_renormalisation_one_loop} we additionally analyse an explicit one-loop example. 

Due to the spectral representation of the sunset graph, the fully perturbative momentum integrals including the subtraction can be solved analytically in both $d=3$ and $d=4$: 

In $d=4$ the sunset graph is superficially quadratically divergent with logarithmic divergences in the sub-diagrams. The respective spectral power counting follows from the momentum dimension of the spectral value, $[\lambda ]=1$, where $[{\cal O} ]$ counts the momentum dimension of $\cal O$. In particular, that of the spectral function is $[\rho(\lambda) ]=-2$. This can be read off from the classical spectral function of a field with mass $m$ with $\rho_\textrm{cl}(\lambda) = 2\pi \delta(\lambda^2-m^2)$. Trivial examples for such a power counting of spectral integrals are \labelcref{eq:SpectralSing} and the propagator itself with $[\int d\lambda\, \lambda \rho(\lambda) )/(\omega^2+\lambda^2)] = -2$. We subtract the zeroth and first order of the Taylor expansion in external momentum about $p^2=\mu^2$ as well as the zeroth term in Taylor expansions about the external momenta of the subdiagrams. These counterterms contribute to the mass renormalisation as well as the wave function renormalisation of the scalar field. The subtractions remove the leading and also the next-to-leading order contributions in the spectral parameters $\lambda_i$ in the integrand of the spectral integral for $\lambda_i\to\infty$. Consequently, the integrand decays faster by two powers of $\lambda_i^2$ and the spectral integrals over $\lambda_i$ with $i=1,2,3$ are UV-finite. 

In $d=3$ there are no divergent subdiagrams and the sunset is superficially logarithmically divergent. We only subtract the zeroth order of the Taylor expansion, which contributes to the mass renormalisation. 
\begin{figure*}[t]
	\centering
	\includegraphics[width=.95\linewidth]{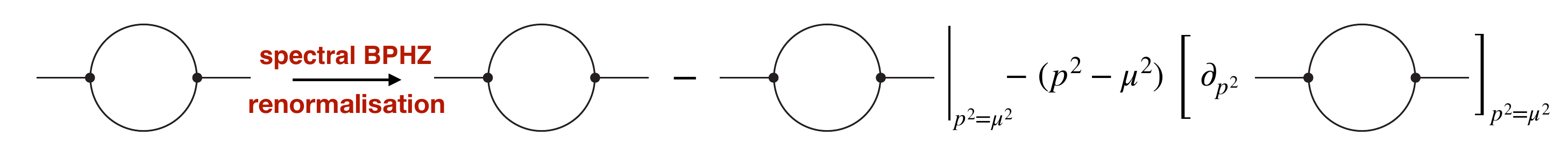}
	\caption{Schematic spectral BPHZ-renormalisation procedure at the example of the one-loop scalar propagator DSE for the $\phi^3$-theory. The diagram is quadratically divergent in $d=6$. First, the loop momentum divergences are discarded by the usual momentum renormalisation part of dimensional regularisation (\ie implicitly assumed on the LHS). By introducing mass and wave function counterterms, the diagram is subtracted by the first two terms of its own Taylor expansion around the RG scale $\mu$. This cancels the leading order quadratic and subleading logarithmic divergences of the spectral integrals.}
	\label{fig:spectral_regularisation_example}
\end{figure*}

We are left with finite spectral integrations for the sunset graph in $d=3,4$ at $\epsilon=0$. The integrand depends analytically on the external momentum, the spectral values $\lambda_i$ with $i=1,2,3$ of the three internal lines as well the respective spectral functions $\rho(\lambda_i)$. In general, the remaining spectral integrals have to be performed numerically.  

In summary, \textit{spectral BPHZ-renormalisation}, as described in detail above, has the same workflow as in the last section and is depicted in \Cref{fig:integration_swap_scheme}: First we apply dimensional regularisation to the momentum integrations and swap the order of momentum and spectral integrals. Then we perform the momentum integration analytically, right bottom corner in \Cref{fig:integration_swap_scheme}. Finally we apply the spectral BPHZ-step: the Taylor expansion in all momenta about the renormalisation scale $\mu$, which allows us to take the limit $\epsilon \to 0$. This leaves us with the task to perform the finite spectral integrals either analytically or numerically, depending on the application.  

\subsection{One-loop example: $\phi^3$-theory in $d=6$} \label{subsubsec:spectral_renormalisation_one_loop}
For further illustration we now apply \textit{spectral BPHZ-renormalisation} within the simple perturbative example of the one-loop DSE for the two-point function of the (renormalisable) $\phi^3$-theory in $d=6$ dimensions with a coupling $g/(3!)\int_x \phi^3$. The advantage of this example is that already at one-loop it requires both a mass renormalisation and a wave function renormalisation. Hence, both can be discussed within a simple one-loop computation. In contrast, in the $\phi^4$-theory the wave function renormalisation only arises at  two-loop from the sunset diagram (even for $\phi_c\neq 0$). 

In the $\phi^3$-theory, the only diagrams in the gap equation \Cref{fig:DSE_2pt} are the polarisation diagram and the squint diagram. At one loop we only have to consider the polarisation diagram. In $d=6$, this diagram is quadratically divergent. The respective DSE for the (inverse) propagator reads schematically, 
\begin{widetext}
\begin{align}
\Gamma^{(2)}(p) = &\; Z_{\phi,0}(\mu)\left(p^2+m_{\phi,0}^2(\mu) \right) \,+ g^2  \int_{\lambda_1,\lambda_2}\lambda_1 \lambda_2 \, \rho(\lambda_1)\rho(\lambda_2)\,F(p,\mu;\lambda_1,\lambda_2) \,.
\label{eq:schematic_scalar_DSE}
\end{align}
The integrand $F$ results from the momentum integration and depends on the renormalisation group scale $\mu$ due to dimensional regularisation. $Z_{\phi,0}$ is the wave function renormalisation and $m_{\phi,0}^2$ is the bare mass squared. Both bare parameters contain counterterms that remove the divergences in the diagrams within an expansion about $p^2=\mu^2$: the constant quadratic divergence as well as the logarithmic divergence proportional to $p^2$. This amounts to  the choices 
\begin{align} \nonumber 
Z_{\phi,0}(\mu) =&\, 1 - g^2 \int_{\lambda_1,\lambda_2}\lambda_1 \lambda_2 \, \rho(\lambda_1)\rho(\lambda_2)\,\left.\frac{F(p,\mu;\lambda_1,\lambda_2)}{\partial{p^2} } \right|_{p^2=\mu^2}\,\,, \\[1ex]
Z_{\phi,0}(\mu)  m_{{\phi,0}}^2(\mu) =&\, m_\phi^2 - g^2 \int_{\lambda_1,\lambda_2}
\lambda_1 \lambda_2 \,\rho(\lambda_1)\rho(\lambda_2) \left[ F(p,\mu;\lambda_1,\lambda_2) - \mu^2 \left.\frac{ F(p,\mu;\lambda_1,\lambda_2)}{\partial{p^2}}\right|_{p^2=\mu^2}\right] \;,
\label{eq:counter_terms}
\end{align}
for wave function and mass renormalisation respectively. The counterterms propertional to $g^2$ in \labelcref{eq:counter_terms} provide the first two terms of the Taylor expansion about $p^2=\mu^2$ of the diagram. To see this, we insert \labelcref{eq:counter_terms} in the DSE \labelcref{eq:schematic_scalar_DSE}, 
\begin{align}
\Gamma^{(2)}(p) = p^2 + m_\phi^2  + g^2  \int_{\lambda_1,\lambda_2}\hspace{-.2cm}\lambda_1 \lambda_2 \, \rho(\lambda_1)\rho(\lambda_2)\, 
\left[ F(p,\mu;\lambda_1,\lambda_2) - F(\mu,\mu;\lambda_1,\lambda_2) - \left(p^2-\mu^2\right) \left. \frac{F(p,\mu;\lambda_1,\lambda_2)}{\partial{p^2}} \right|_{p^2=\mu^2} \right] \,.
\label{eq:renormalized_scalar_DSE}
\end{align}
\end{widetext}
Eq.~\labelcref{eq:renormalized_scalar_DSE} is depicted in \Cref{fig:spectral_regularisation_example}. Accordingly, \labelcref{eq:counter_terms} implements the standard renormalisation conditions, that the quantum corrections vanish at $p^2=\mu^2$, 
\begin{align} \nonumber 
	\Gamma^{(2)}(p^2=\mu^2) \; = &\; Z_\phi(\mu^2+m_\phi^2) \\[1ex]
	\partial_{p^2}\Gamma^{(2)}(p^2=\mu^2) \; = &\; Z_\phi \,,
\label{eq:rg_conditions_sc}
\end{align}
for the two-point function with $Z_\phi=1$. In the present $\phi^3$-example these two renormalisation conditions are complemented by that for the coupling $g$, which is also logarithmically divergent in $d=6$. As we have introduced this example only for illustration of spectral BPHZ-renormalisation, we refrain from discussing this any further. Mode details on the spectral renormalisation conditions can be found in \Cref{subsec:calculation_analytic_continuation}.

Continuing with the discussion of spectral renormalisation for the two-point function, the subtraction of $F$ in~\labelcref{eq:renormalized_scalar_DSE} by its own Taylor expansion at $p^2=\mu^2$ leads to finite spectral integrals. We emphasise that this is not achieved by simply subtracting the $1/\epsilon$-terms before performing the spectral integration: Since $F$ scales with $\lambda^2$ for large $\lambda$ with $\lambda=\lambda_1,\lambda_2$, the spectral integrals in~\labelcref{eq:schematic_scalar_DSE} are quadratically divergent. After \textit{spectral BPHZ-renormalisation} however, the subtracted scalar integrand  in~\labelcref{eq:renormalized_scalar_DSE} scales as $1/\lambda^{2}$. The subtraction scheme cancels the leading and subleading contributions in $\lambda$ to $F$ and leads to finite spectral integrals. 

\subsection{Non-perturbative spectral renormalisation}
\label{subsec:spectral_renormalisation_beyond_one_loop}
The discussions of the last three sections, \Cref{subsec:spectral_renormalisation} to \Cref{subsubsec:spectral_renormalisation_one_loop}, entail that spectral renormalisation leads to two different parts in the counterterms: the first part is related to the momentum divergences and has all the properties of the counterterms in dimensional regularisation. The second part comes from the spectral divergences. The counterterms in spectral dimensional renormalisation respects all symmetries including gauge symmetries and is tantamount to dimensional regularisation and renormalisation. The counterterms in spectral BPHZ-renormalisation lack the full symmetries. In particular in gauge theories the BPHZ-counter terms are necessarily not gauge-invariant, precisely for restoring gauge consistency of the full renormalised result.   

Importantly, in both spectral renormalisation schemes, the spectral counterterms follow the same recursive relations known from standard perturbation theory. This makes it a consistent renormalisation scheme to all orders of perturbation theory. 

In non-perturbative applications of the present spectral approach, the non-perturbative information is solely present in the spectral functions of propagators and vertices. Moreover, within the DSE we only deal with one- and two-loop diagrams with $n$ vertices derived from the master DSE in \labelcref{eq:DSE}, see also \Cref{fig:DSE_1pt}.  We have one classical (bare) vertex and $n-1$ full vertices as well as full propagators.  If recursively written in terms of loops, both, the \textit{finite} full vertices and propagators, carry subtractions of the divergences in these diagrams that render the diagrams finite. The leftover subtractions from these re-distribution renormalise the one or two explicit loops in the DSE diagrams.

In summary, non-perturbative spectral renormalisation only concerns the counterterms for the explicit loops in the DSE, while the rest of the renormalisation is carried by the finite full vertices and propagators that have to obey the renormalisation conditions. This is a consistent numerical non-perturbative renormalisation scheme. 
\section{\texorpdfstring{$\phi^4$-theory}\ \ in 2+1 dimensions}
\label{sec:Calculation}
The $\phi^4$-theory in $2+1$ dimensions is super-renormalisable, and the initial two renormalisation conditions for the two-point function in \Cref{fig:DSE_2pt} reduce to the first one for the mass. Moreover, we do not need to renormalise the coupling. This entails that in the DSE for the scalar two-point function spectral BPHZ-renormalisation as discussed in \Cref{subsec:SpecBPHZ} simply amounts to subtracting the zeroth order term in the Taylor expansion about $p^2=\mu^2$. After the momentum integrals are computed analytically within dimensional regularisation, we are left with the finite spectral integrals. 

After completing renormalisation, the iterative solution procedure in the DSE is briefly described as follows: With a given input spectral function the renormalised DSE is evaluated in Minkowski space-time. The input spectral function is either the initial guess or the result of the last iteration step. Then, an updated retarded two-point function is computed from the result. This allows us to extract an updated spectral function, which is fed back as input into the next iteration step. In this section we discuss the calculation sketched above in a detailed way, step by step. 

\subsection{Momentum integration and spectral renormalisation} \label{subsec:calculation_momentum_integration}
The DSE can be expressed as the sum of the bare two-point function and the loop diagrams $D_j$,  
\begin{align}
\label{eq:DSE_eucl_general}
	\Gamma^{(2)}(p) = p^2 + m^2_{\phi,0} + \sum_{\{j\}} D_j(p) \;,
\end{align}
with $j = \text{tad,pol,sun,squint}$. In \labelcref{eq:DSE_eucl_general} we have used that the $\phi^4$-theory in $d=3$ is super-renormalisable and the only divergent term is the mass term. From now on, we drop the $\mu$-dependence of the spectral integrands for notational simplicity. With the K{\"a}llen-Lehmann spectral representation for the full propagator~\labelcref{eq:kaellenlehmann}, as well as momentum-independent vertices, an arbitrary loop diagram $D_j$ in the DSE takes the form, 
\begin{equation}\label{eq:DSE_eucl}
D_j(p) = g_j \prod_{i}^{N_j} \left( \int_{\lambda_i} \lambda_i \rho(\lambda_i) \right) \; I_j(p;\lambda_1, ...,\lambda_{N_j}) \,.
\end{equation}
The prefactors $g_j$ are the products of the combinatorial prefactors in the DSE and the vertices of the corresponding diagram. In \Cref{tab:prefactors_DSE} we provide the prefactors for the Minkowski version of \labelcref{eq:DSE_eucl}, see \labelcref{eq:DSE_mink}. 

$D_j$ has $N_j$ internal lines, each of them coming with one spectral integral and a corresponding spectral function. The $I_j$ are nothing but a product of (momentum) loop integrals over $N_j$ classical propagators with different spectral masses $\lambda_i$,
\begin{equation}
\label{eq:momentum_integrals_schematic}
	I_j(p;\lambda_1, \ldots ,\lambda_{N_j}) = \prod_{k}^{N_j^{\mathrm{loops}}} \int \frac{d^3q_k}{(2\pi)^3}  \prod_{i}^{N_j} \frac{1}{\lambda_i^2 + l_i^2}\,. 
\end{equation}
In \labelcref{eq:momentum_integrals_schematic} the momenta $l_i$ are linear combinations of the loop-momenta $q_k$ and the external momentum $p$. The number of loops is denoted by $N_j^{\mathrm{loops}}$ in \labelcref{eq:momentum_integrals_schematic}. The analytic solutions of these integrals in $d=3$ are known from perturbation theory, e.g.~\cite{Rajantie:1996np}, and can be used here.

With the analytic expressions for $I_j$, we apply spectral BPHZ-renormalisation, c.f.~\Cref{subsec:SpecBPHZ}. Please note that only the logarithmically divergent sunset diagram $D_{\text{sun}}$ explicitly requires renormalisation, since the bare tadpole simply can be absorbed in the definition of the bare mass squared. To cancel the logarithmic divergence of this diagram, within the BPHZ-approach we subtract the zeroth order term in the Taylor expansion of $D_{\text{sun}}$ about $p^2 = \mu^2$. The renormalised diagram reads 
\begin{align} \label{eq:sunset_renormalized} \nonumber
D_{\text{sun}}^{\text{ren}}(\omega) = \; &  g_{\text{sun}} \int_{\lambda_1,\lambda_2}  \lambda_1 \lambda_2 \rho(\lambda_1) \rho(\lambda_2)  \\[1ex] 
& \hspace{.4cm} \times \big[ I_{\text{sun}}(\omega;\lambda_1,\lambda_2) - I_{\text{sun}}(\mu;\lambda_1,\lambda_2) \big] \,.
\end{align}
%
\subsection{Analytic continuation} \label{subsec:calculation_analytic_continuation}

The $I_j(p)$ are evaluated in Minkowski space-time for the retarded two-point function. This is done by parametrising the complex (Euclidean) frequency as $p_0=-\text{i} (\omega+\text{i} \varepsilon)$ and taking the limit $\varepsilon \to 0$. In a slight abuse of notation we denote the continued expression as $I_j(\omega)$. They are given explicitly in \Cref{app:analytic_expressions}. The DSE in Minkowski space-time reads 
\begin{align}\nonumber 
 	\Gamma^{(2)}(\omega) = &\; -\omega^2 + m^2_{\phi,0} \\[1ex] 
 	& \hspace{-1cm} +\sum_{\{j\}} g_j \prod_{i}^{N_j} \left( \int_{\lambda_i} \lambda_i \rho(\lambda_i) \right)  I_j(\omega,\mu;\lambda_1, ...,\lambda_{N_j}) 
 	\,.
 \label{eq:DSE_mink}	
\end{align}
The prefactors of the diagrams of the DSE \labelcref{eq:DSE_mink}  with the classical vertex approximation and the skeleton expansion can be found in \Cref{tab:prefactors_DSE}. It can be deduced from the analytic structure of the spectral integrands $I_j(\omega)$, that the support in $\omega$ of the imaginary part of the polarisation diagram starts at $\omega = \lambda_1 + \lambda_2$, as expected. Similarly, we find that the support of the sunset diagram starts at $\omega = \lambda_1 + \lambda_2 + \lambda_3$. Hence, the support of the expressions in the two-point function is given by multiples of the pole mass. This shows how the Cutkosky cutting rules can be easily extracted from the present spectral approach. 

It is clear from \labelcref{eq:DSE_mink} that the \textit{spectral renormalisation} approach allows for the implementation of physical on-shell renormalisation conditions. This is in contrast to Euclidean computations, where on-shell renormalisation can only be implemented for massless modes. On-shell renormalisation has the advantage that it minimises the quantum corrections in a study of 
the resonance spectrum of a given theory. The renormalisation conditions  \labelcref{eq:rg_conditions_sc} for a $\phi^4$-theory in $d=3$ dimensions reduce to 
\begin{equation} \label{eq:RGOnshell}
\Gamma^{(2)}[p^2=-m_\textrm{pole}^2]= \,0  \,, 
\end{equation}
since both the coupling and the kinetic term do not require renormalisation. The triviality of the wave function renormalisation or rather the vanishing anomalous dimension entail, that the leading large momentum behaviour of the propagator is given by $1/p^2$. Accordingly, the canonical commutation relations of the scalar field are unchanged, as is the normalisation of dynamical states. The sum rule~\labelcref{eq:scalarsumrule} is hence formally always satisfied in $2+1$ dimensions, and thus provides a non-trivially benchmark test of our results. 

\subsection{Spectral integration and iteration} \label{subsec:spec_integration_iteration}
\begin{figure}[t] 
	\centering
	\hspace{80pt}
	\includegraphics[width=\linewidth]{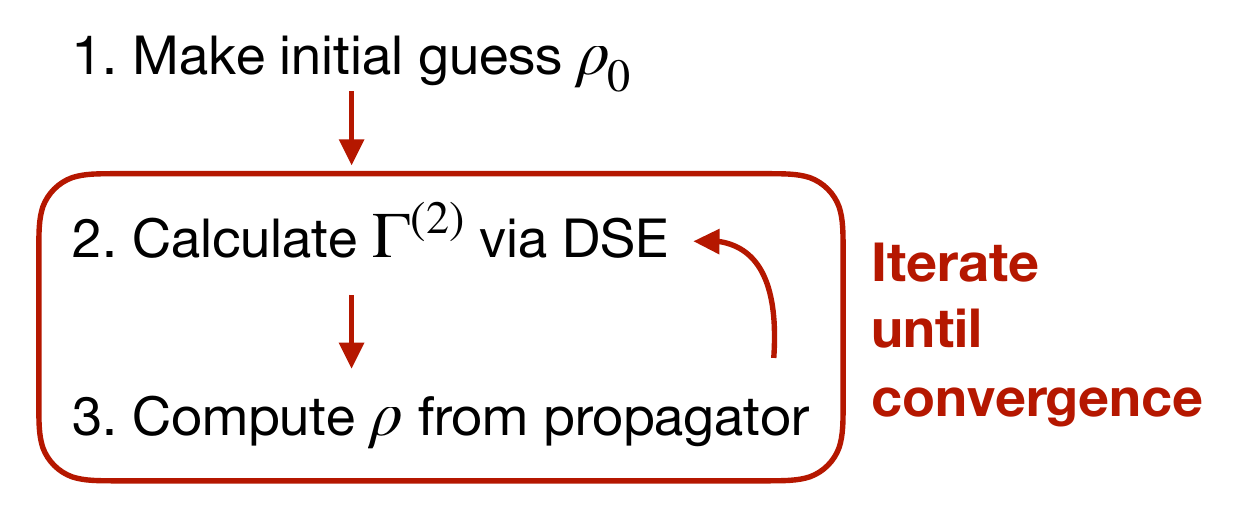}
	\caption{Iteration procedure for computing the spectral function. With the initial guess $\rho_0$ for the spectral function, the two-point function $\Gamma^{(2)}$ is computed via the DSE. The resulting spectral function is fed back into the DSE for the two-point function. This procedure is iterated until the convergence for the spectral function is reached.}
	\label{fig:iteration_scheme}
\end{figure}
\begin{figure*}[t]
	\centering
	\includegraphics[width=\linewidth]{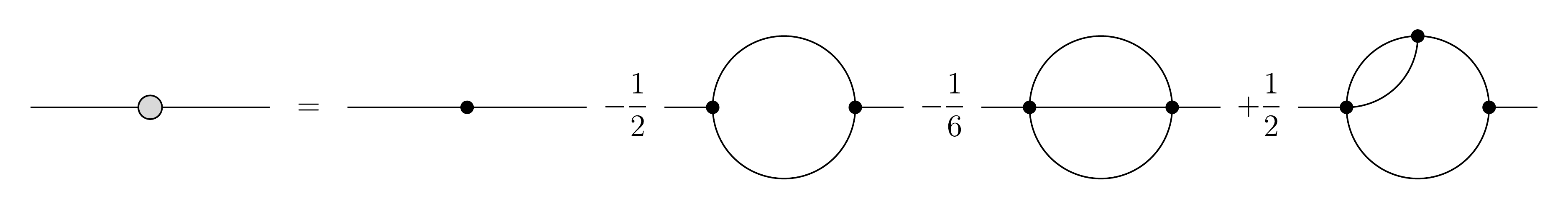}
	\caption{DSE of the two-point function with the classical vertex approximation as used in \Cref{subsec:results_bare_vertices}. The tadpole is not present because its contribution is absorbed into the mass renormalization in the bare inverse propagator, for more details can be found in the main text. The two-loop terms constitute vertex corrections of the classical vertices in the one-loop diagrams. The DSE is not two-loop complete as further vertex corrections have been dropped due to the classical vertex approximation. The notation is given in \Cref{fig:FunctionalMethods:DSE_notation}.}
	\label{fig:DSE_2pt_bare}
\end{figure*}
The remaining multi-dimensional integrals over the spectral parameters $\lambda_i$ in~\labelcref{eq:DSE_mink} are solved numerically. Details on the numerical part of the calculation can be found in \Cref{app:Technical_details}. Subsequently, the spectral function is extracted from the updated two-point function via~\labelcref{eq:spec_func_def} and fed back into the DSE. In this way, the DSE is solved iteratively by successively integrating the right hand side of the DSE with the updated spectral function from the last step until the solution converges, comp. \Cref{fig:iteration_scheme}. Note that all the dynamical information is stored in the spectral functions, and the integrands $I_j$ do not change within the iterations. Hence, each iteration only involves the numerical solution of the respective multidimensional spectral integral of each diagram. 

More details on the convergence test can be found in \Cref{app:Technical_details}. There, the rapid convergence is illustrated at an exemplary case, see in particular \Cref{fig:specFuncConvPlot}. 

As a starting point for the iterative procedure, an initial guess for the spectral function has to be made that is close enough to the solution (in the attraction basin of the solution in terms of the iteration). In the present case with the on-shell renormalisation \labelcref{eq:RGOnshell} the spectral function of the classical theory carries already the correct pole position by definition. In general this will improve the convergence properties of the iteration.  This is yet another property that singles out on-shell renormalisation. The classical spectral function is given by, 
\begin{align} \nonumber 
\rho_0(\omega) = &\, \delta(\omega^2-m_{\text{pole}}^2) \\
= & \, \frac{\pi}{ m_{\text{pole}}} \big[ \delta (\omega - m_{\text{pole}}) - \delta(\omega + m_{\text{pole}}) \big] \,.
\label{eq:initial_guess_spec_func}
\end{align}
The delta-function peaks are located at the physical pole mass squared, and the delta-functions at $ \pm m_{\text{pole}}$, are related by anti-symmetry.

The branch cuts of the loop corrections generate a continuous tail in the spectral function when inserting the initial spectral function $\rho_0$ on the right hand side of the DSE. This entails via the sum rule  in~\labelcref{eq:scalarsumrule}, that the residue of the mass pole decreases to $Z<1$ due to the positive weight of the tail. The mass pole residue of the updated spectral function is obtained via the relation
\begin{equation} \label{eq:residue_mass_pole}
Z= - \frac{2 m_{\text{pole}}}{\partial_{\omega}\Gamma^{(2)}(\omega)}\bigg|_{\omega=m_{\text{pole}}}
\,.
\end{equation}
The counterterm for the mass is conveniently extracted from $\text{Re} \; \Gamma^{(2)}(\omega= m_{\text{pole}})=0$. 

The three-point function $\Gamma^{(3)}[\phi_c]$ in the gap equation is evaluated at the constant field value $\phi_c = \phi_0$, which solves the equation of motion, $\partial_\phi V_{\text{eff}}[\phi_0]=0$, at each step of the iteration. For fields in the vicinity of $\phi_0$ we can expand the effective potential in powers of $\phi^2-\phi_0^2$ in the broken phase, leading to,   
\begin{align}
\label{eq:effective_potential}
V_{\text{eff}}[\phi] = \sum_{n=2}^\infty \frac{v_n}{2n!} \left(\phi^2 - \phi_0^2\right)^n \,.
\end{align}
Accordingly, the two-, three- and four-point functions at vanishing momentum are given by, 
\begin{align}\nonumber 
\Gamma^{(2)}[\phi_0]=&\,\frac{1}{3} v_2\,\phi_0^2 \,,\\[1ex]\nonumber  
 \Gamma^{(3)}[\phi_0]= &\,v_2\, \phi_0\left(1 +\frac{1}{15} \frac{v_3}{v_2} \phi_0^2\right)\,,\\[1ex]
 \Gamma^{(4)}[\phi_0]=&\, v_2 \left( 1 +\frac{1}{5}\frac{v_3}{v_2}\phi_0^2+ \frac{1}{105} \frac{v_4}{v_2}\phi_0^4\right) \,. 
\label{eq:Vertices_full} 
\end{align} 
Dropping the higher order terms $O\big((\phi^2-\phi_0^2)^3\big)$ in \labelcref{eq:effective_potential}, the terms in parentheses of~\labelcref{eq:Vertices_full} all reduce to unity. We introduce the curvature mass $m_\mathrm{cur}^2(m_\mathrm{pole},\lambda_\phi)$ as the value of the two-point function at vanishing momentum, i.e., 
 \begin{equation}\label{eq:mcur}
 m_\mathrm{cur}^2(m_\mathrm{pole},\lambda_\phi) \equiv \Gamma^{(2)}[\phi_0](p=0) \,.
 \end{equation} 
 Note that $m_\mathrm{cur}$ is determined by the two free parameters of the theory, $m_\mathrm{pole}$ and $\lambda_\phi$. It is no new parameter. Using \labelcref{eq:mcur} in \labelcref{eq:Vertices_full} and  dropping the higher-order contributions, the three- and four-point functions at vanishing momentum are given by, 
\begin{align}\nonumber 
\Gamma^{(3)}[\phi_0]= &\, \sqrt{3 \; v_2}\, m_\mathrm{cur} \,,\\[1ex]
\Gamma^{(4)}[\phi_0]=&\, v_2 \,. 
\label{eq:Vertices}  
\end{align} 
$v_2$ is nothing but the full four-vertex at vanishing momentum,  $v_2=\Gamma^{(4)}[\phi_0](p=0)$. For a general, momentum dependent four-point function in the $s$-channel, we find for the momentum-dependent three-point function, 
\begin{align}
\label{eq:three_vert}
\Gamma^{(3)}(p) = \ \Gamma^{(4)}(p) \sqrt{\frac{3 \;  }{v_2}} m_\mathrm{cur}\,,
\end{align}
dropping the $\phi_0$-dependence of the correlators from now on. Note that by choosing $\Gamma^{(3)}$ consistently with $\Gamma^{(4)}$ as done above, the three-point function becomes dynamical and is hence updated through each iteration by its dependence on the two- and four-point function.

\section{Results}
\label{sec:results}

In this section we compute and discuss the solution to the DSE for the propagator with two different approximations for the vertices. At first we solve the DSE with the classical four-point vertex and 
the related three-point vertex as derived in \Cref{subsec:spec_integration_iteration}. Subsequently, the DSE is considered in a skeleton expansion. Both, the full three- and four-point functions $\Gamma^{(3)}, \Gamma^{(4)}$, are based on the bubble resummed $s$-channel approximation to the four-point vertex derived from its Bethe-Salpeter equation. We also  use the DSE  for $\Gamma^{(4)}$ to compute results within a self-consistent version of this setup.

\subsection{DSE with classical four-point vertex} \label{subsec:results_bare_vertices}

\begin{figure*}[t]
	\centering
	\includegraphics[width=.48\linewidth]{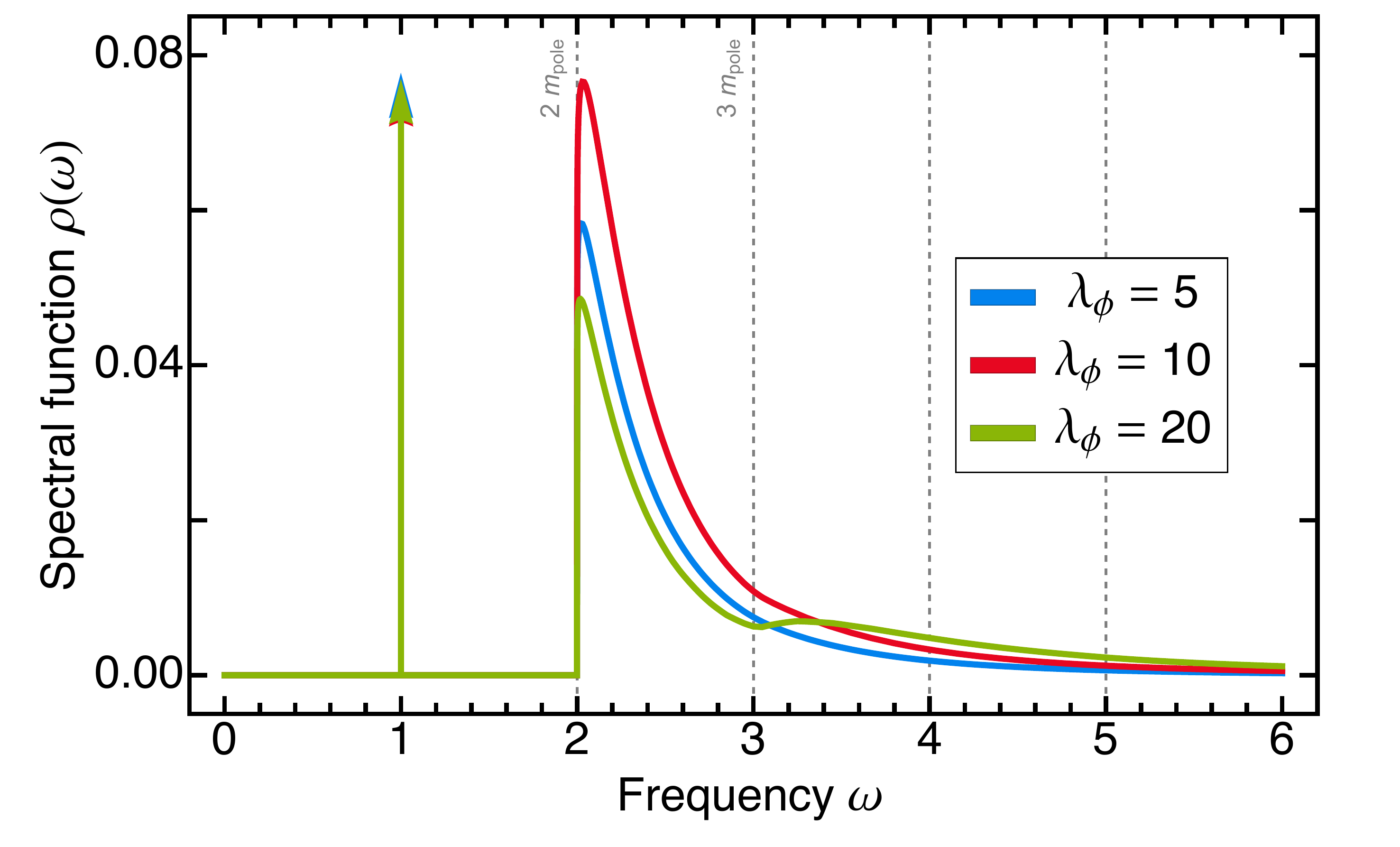}
	\includegraphics[width=.48\linewidth]{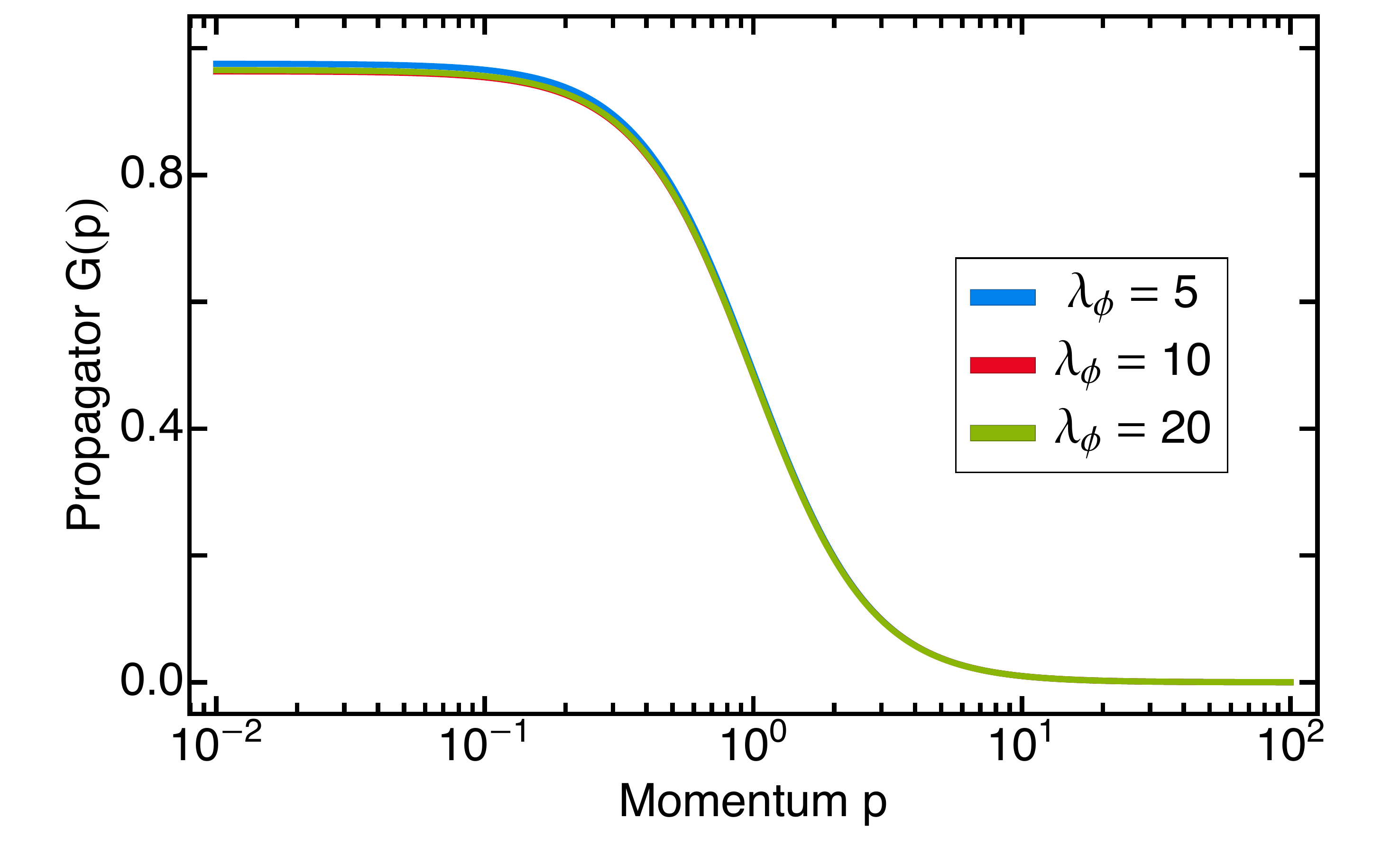}
	\caption{Spectral function (left) and propagator (right) in the scalar theory for the coupling choices ${\lambda}_{\phi} =5,10,20$ from the full DSE with classical vertices using on-shell renormalisation~\labelcref{eq:RGOnshell}. All dimensionful quantities were rescaled in units of the respective mass pole result. The different height of the delta peaks encodes the magnitude of the residue relative to the other spectral functions. The grey dashed lines mark the n-particle onsets. For large enough coupling, the three-particle onset becomes visible. The propagators were computed by the K\"allen-Lehmann spectral representation. }
	\label{fig:scalar_results_rescaled}
\end{figure*}

In the present section we approximate the full vertices in the gap equation in \Cref{fig:DSE_2pt} with their classical counterparts while keeping the two-loop terms. This is depicted in \Cref{fig:DSE_2pt_bare}. The two-loop terms constitute vertex corrections to the classical four-point function in the tadpole diagram (sunset) and for the classical three-point function in the polarisation diagram (squint). In the latter case this is but half of the vertex correction, the other half has been dropped in the current approximation when approximating the full three-point function in the polarisation diagram by its classical counterpart. This approximation resums the propagator but is expected to fail in a regime where vertex corrections grow large. 

For the four-point function the classical vertex approximation amounts to,  
\begin{align}
\label{eq:vertex_approx_sc}
	\Gamma^{(4)} =  S^{(4)} =\lambda_\phi
\,.
\end{align}
With \labelcref{eq:vertex_approx_sc} and \labelcref{eq:Vertices}, the three-point function is given by 
\begin{align}\label{eq:classGamma3}
\Gamma^{(3)} = S^{(3)}[\phi_0]= \sqrt{ 3\lambda_\phi} m_{\text{cur}} \,, 
\end{align} 
see \labelcref{eq:Vertices}. The resulting prefactors $g_j$ of the diagrams are listed in \Cref{tab:prefactors_DSE}. The tadpole diagram only contributes a momentum-independent term that shifts the mass, and is absorbed into the mass renormalisation. 

All units will be given in the value of the pole mass. Put differently, we introduce dimensionless units 
\begin{align}
 \lambda_\phi\to  \frac{\lambda_\phi}{m_\textrm{pole}}\,, \quad \omega \to 
 \frac{\omega}{m_\textrm{pole}}\,,\quad p \to 
 \frac{p}{m_\textrm{pole}}\,, 
\label{eq:mpoleUnits}\end{align}
This also entails $m_\textrm{pole}=1$, and the massless limit is taken with $\lambda_\phi\to\infty$. The dimensionless units also emphasise the well-known fact, that the physics of the $\phi^4$-theory in $d=2+1$ is specified by one dimensionless parameter, the ratio of coupling and (pole) mass. 

We solve the DSE for three different values of the classical coupling constant ${\lambda}_\phi = 5, 10, 20$. The renormalised mass is fixed by the on-shell RG-conditions~\labelcref{eq:RGOnshell} with  $m_\text{pole} = 1$, and the quantum corrections to the mass vanish on-shell. For the smallest classical coupling used here,  ${\lambda}_\phi = 5$, we take the classical spectral function as initial choice. For the further couplings we use as the initial choice the full quantum spectral function of the closest coupling value available. This stabilises the iterative procedure when successively moving further to larger couplings inducing larger quantum corrections.

The resulting spectral functions, as well as the corresponding propagators, are shown in \Cref{fig:scalar_results_rescaled} for different values of the classical four-point coupling, $\lambda_\phi=5,10, 20$. The mass pole as well as the onset of the two-particle threshold, $\phi \to \phi \phi$, at twice the pole mass are clearly visible. The three-particle threshold, $\phi \to \phi \phi \phi$, at $3m_{\textrm{pole}}$ becomes visible for large enough coupling. Also, all higher $n$-particle thresholds are present in the result. Since they are suppressed by the inverse of the respective threshold energy squared, they are not visible in the plots. The main effect of a stronger coupling (or small pole mass) can be understood intuitively very well. The residue of the mass pole becomes smaller, while the scattering cut gets larger contributions, i.e. scatterings get enhanced due to the large coupling or the small pole mass. As mentioned before, the only parameter present is $\lambda_\phi/m_\textrm{pole}$, and in the present units with $m_\textrm{pole}=1$ this simply is $\lambda_\phi$, see \labelcref{eq:mpoleUnits}. At the largest coupling value considered here, $\lambda_\phi=20$, also the higher scattering processes start contributing significantly: the threshold at $3 m_\textrm{pole}$ is clearly visible in the spectral function for this coupling,  shown in the left panel of \Cref{fig:scalar_results_rescaled}. For large couplings also the higher thresholds kick in.  This scattering physics also leaves its traces in the Euclidean propagator, shown in the right panel of \Cref{fig:scalar_results_rescaled}: while first the propagator drops for small momenta with the increasing coupling (measured in the respective pole masses), it increases again for even larger ones due to the more pronounced scattering physics present in the spectral function. 

\subsection{Fully non-perturbative DSE} \label{subsec:res4ptfct}

\begin{figure*}[t]
	\centering
	\includegraphics[width=\linewidth]{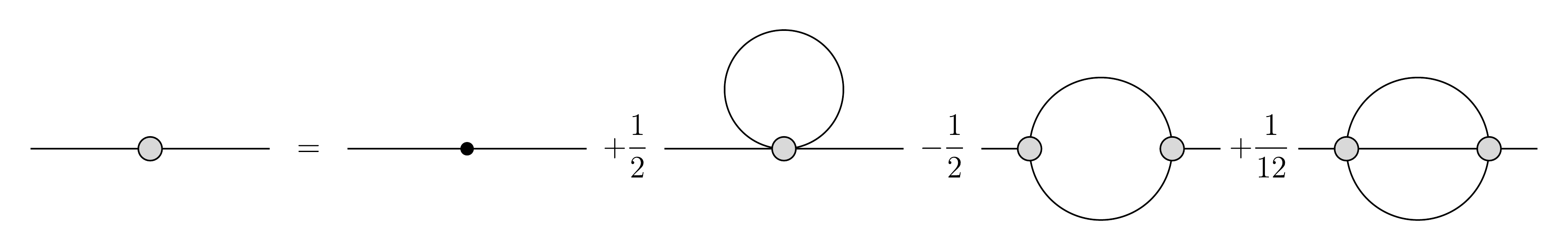}
	\caption{Truncated DSE of the two-point function as used in \Cref{subsec:solution_coupled_system}. Notation is given in \Cref{fig:FunctionalMethods:DSE_notation}. The full $s$-channel four-vertex in the tadpole diagram enters via its spectral representation~\labelcref{eq:spec_rep_vert}. The full vertices in polarisation and sunset are approximated at zero frequency, see~\Cref{subsubsec:vertex_approximation_skeleton}. The different prefactor in front of the sunset diagram as compared to~\Cref{fig:DSE_2pt_bare} is due to the tadpoles contribution to the the sunset topology, cf.~\Cref{subsubsec:tadpole_contribution_sunset}. Squint, kite and double-bubble topology are dropped, as motivated in~\Cref{subsec:res4ptfct}.}
	\label{fig:DSE_2pt_full}
\end{figure*}

The practical applicability of the spectral renormalisation scheme has been shown in the last section within the DSE for the two-point function with classical vertices, see \Cref{fig:DSE_2pt_bare}. This approximation implements a full  resummation of the propagator. The approximation also includes some corrections in the higher order diagrams in the DSE as already discussed in the last section. While these diagrams contribute to the higher-order scattering thresholds, this may not be sufficient in the limit of asymptotically large couplings $\lambda_\phi/m_\textrm{pole}\to \infty$, also tantamount to small pole masses. In this regime  the vertex corrections should be taken into account  consistently. 

In the present section we discuss non-perturbative expansion schemes of the DSE as well as resummations of the vertices. This allows us to study the strongly correlated regime of the theory. A full quantitative study is beyond the scope of the present contribution and is deferred to future work. 

A non-perturbative expansion scheme for the DSE is given by the skeleton expansion. In this expansion all vertices are full-dressed, and higher loop-order diagrams with dressed propagators and vertices have to be introduced successively. Instead of an expansion in classical vertices it is an expansion in fully dressed ones. This expansion is closely related to nPI-resummation schemes, in the $\phi^4$-theory it is related to a 4-PI scheme. 

Here we consider the two-loop order of the skeleton expansion in the broken phase. A first observation is, that the prefactor of the squint diagram vanishes: it is fully contained in the polarisation diagram with two dressed three-point functions $\Gamma^{(3)}$. At perturbative two-loop level the expansion involves a kite-diagram as well as a double-bubble diagram. Both topologies are only generated from the polarisation diagram in the DSE, more precisely from vertex corrections of the dressed three-point function. These contributions have to be subtracted in terms of explicit kite and double-bubble diagram in the skeleton expansion. For small field expectation values $\phi_0 \ll 1$, it is reasonable to neglect the kite diagram, since it scales like $\phi_0^4$ due to its four three-point functions. We will also drop the double-bubble diagram which is of order $\phi_0^2$. These approximations are discussed again later. The remaining diagrams are the polarisation, sunset and tadpole. The present approximation of the two-loop skeleton expansion of the gap equation is depicted in depicted in \Cref{fig:DSE_2pt_full}.

\begin{figure*}[t]
	\centering
	\includegraphics[width=.9\linewidth]{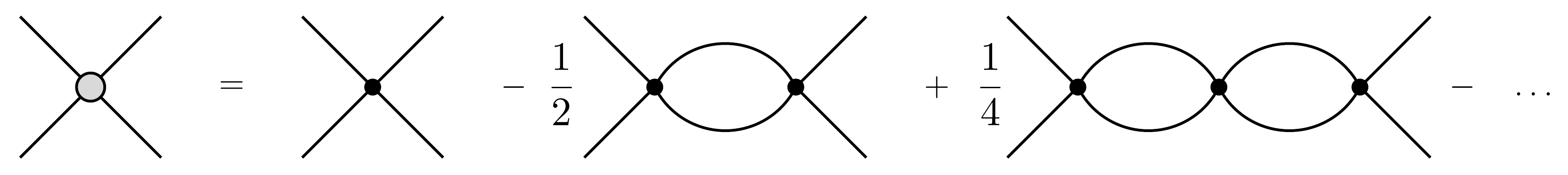}
	\caption{$s$-channel expansion of the momentum dependent four-point function $\Gamma^{(4)}$. With a bubble resummation one arrives at~\labelcref{eq:bubble_resummation}. The notation is defined in \Cref{fig:FunctionalMethods:DSE_notation}. }
	\label{fig:bubble_resummation_diagr}
\end{figure*}
\subsubsection{Bubble-resummed s-channel four-point function} \label{subsubsec:bubble_resummation}
It is left to specify the approximations for the three-point and four-point vertices. To begin with, we still use the relation \labelcref{eq:three_vert} for the three-point function with the assumption of small field values $\phi_0^2$. This leaves us with the four-point function, for which we resort to a bubble-resummed $s$-channel expansion, e.g.~\cite{Peskin:1995ev,Fukushima:2012xw}, shown graphically in \Cref{fig:bubble_resummation_diagr}. Algebraically, the momentum dependence of the $s$-channel in the four-point function can be expressed as 
\begin{equation} \label{eq:bubble_resummation}
\Gamma^{(4)}(p) \; = \; \frac{\lambda_\phi}{1+ \lambda_\phi \Pi_{\text{fish}}(p)} \,.
\end{equation}
Here, $\Pi_{\text{fish}}$ is the one-loop part of the $s$-channel self-energy (apart from $\lambda_\phi^2$),  
\begin{equation} \label{eq:self_energy}
\Pi_{\text{fish}}(p) \; = \; \frac{1}{2} \int_{\lambda_1,\lambda_2} \!\!\!\! \lambda_1 \lambda_2 \rho(\lambda_1) \rho(\lambda_2)  \; I_{\text{pol}}(p;\lambda_1,\lambda_{2}) \,. 
\end{equation}
This is exactly the polarisation diagram also appearing in the DSE with prefactor $g_{\text{pol}} = 1$, \cf  \Cref{fig:spectral_regularisation_example} and \labelcref{eq:schematic_scalar_DSE}. The analytically-continued expression for $\Gamma^{(4)}(\omega)$ is obtained by simply replacing $I_{\text{pol}}(p)$ in~\labelcref{eq:self_energy} with $I_{\text{pol}}(\omega)$. 

The resulting $\Gamma^{(4)}(\omega)$ depends on the full propagator through the spectral functions in~\labelcref{eq:self_energy}, and the equations for $\Gamma^{(2)}$ and $\Gamma^{(4)}$ are coupled. We also note that the $s$-channel resummation used here is obtained in the NLO-expansion of the $1/N$-expansion of an O($N$)-theory with $N$ real scalar fields. 

The iteration procedure does not change for such a coupled system, even though coupled systems generically show worse convergence properties. For a given input pair $\Gamma^{(2)}$ and $\Gamma^{(4)}$ we compute the next iteration from the right hand side of the DSE for $\Gamma^{(2)}$, \Cref{fig:DSE_2pt_full}, and the resummed representation of $\Gamma^{(4)}$, \labelcref{eq:bubble_resummation}. This is repeated until convergence is reached.

\subsubsection{Tadpole contribution to the sunset topology} \label{subsubsec:tadpole_contribution_sunset}
Before turning to the explicit approximation used in the skeleton scheme, we emphasise again, that the fully-dressed tadpole diagram and the fully-dressed sunset diagram are related. They both carry the $s$-channel of the four-point vertex. While the tadpole simply is proportional to the $s$-channel four-point vertex, the sunset includes the fish-diagram as a sub-diagram.  Indeed, the perturbative two-loop sunset graph is a combination of the respective contributions, and the prefactor $g_{\text{sun}}$ of the sunset diagram in the skeleton expansion is such that the perturbative prefactor , c.f.~\Cref{tab:prefactors_DSE}. 

The tadpole diagram is proportional to the $s$-channel four-point vertex and we use the full momentum-dependent four-point vertex obtained from the bubble resummation~\labelcref{eq:bubble_resummation}. Inserting the diagrammatic vertex expansion explicitly into the diagram, one sees that the dressed tadpole contributes to the sunset topology on the perturbative two-loop level. However, this contribution does not account for the full prefactor of the latter.  To arrive at the correct perturbative prefactor of the sunset diagram, the prefactor of the fully-dressed sunset diagram in the skeleton expansion needs to be adjusted accordingly, see \Cref{fig:DSE_2pt_full} and \Cref{tab:prefactors_DSE}. 

\subsubsection{Vertex approximation in the skeleton expansion} \label{subsubsec:vertex_approximation_skeleton}
In the sunset diagram, the two four-point vertices are averaged due to the two loop momenta that run through both vertices. This averaging holds true for both the Euclidean branch as well as the Minkowski one. For this reason we approximate the full momentum-dependent four-point vertices by that at vanishing momentum, $\Gamma^{(4)}[\phi_0]$. 

\begin{figure*}[t]
	\centering
	\includegraphics[width=.48\linewidth]{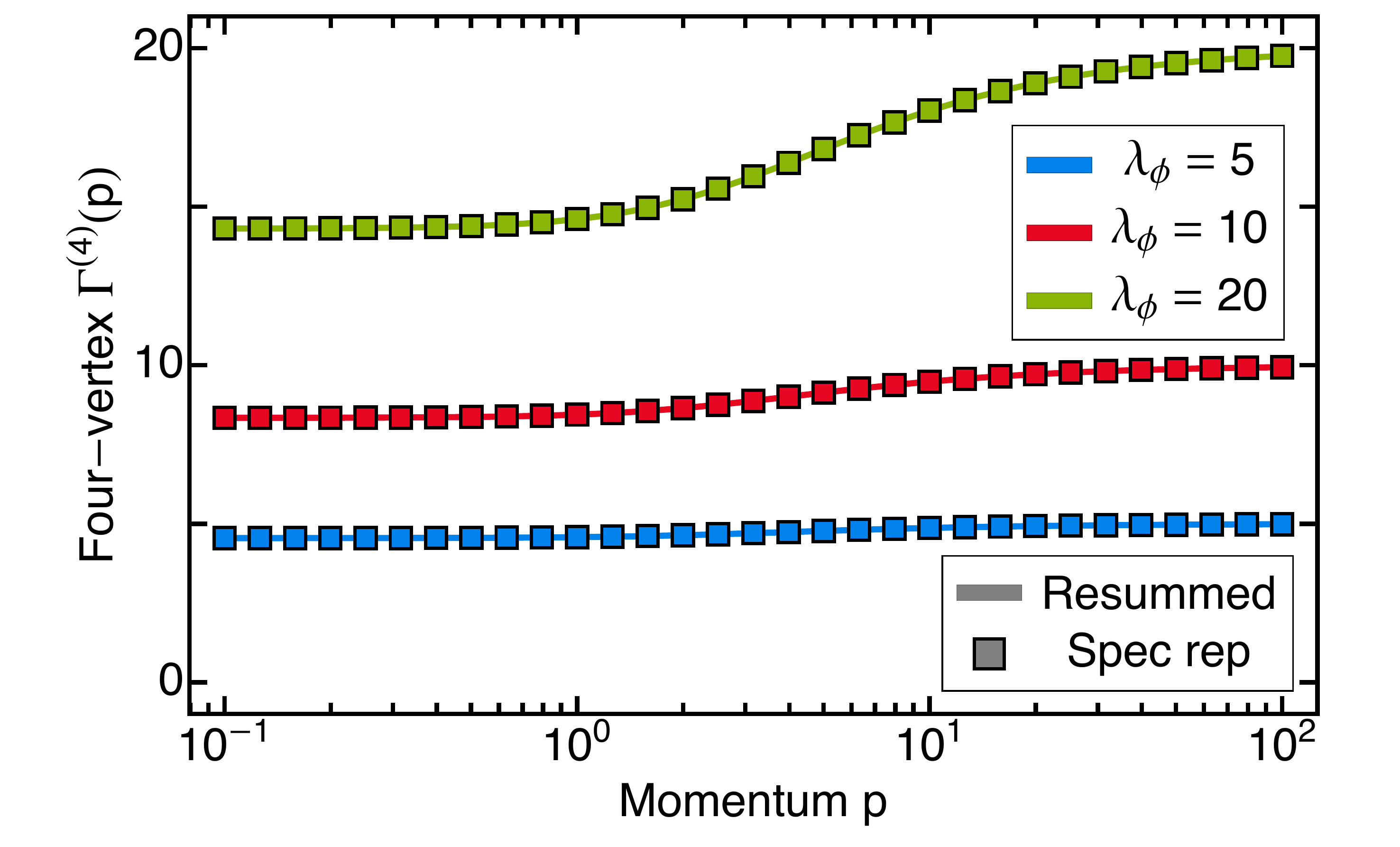}
	\includegraphics[width=.48\linewidth]{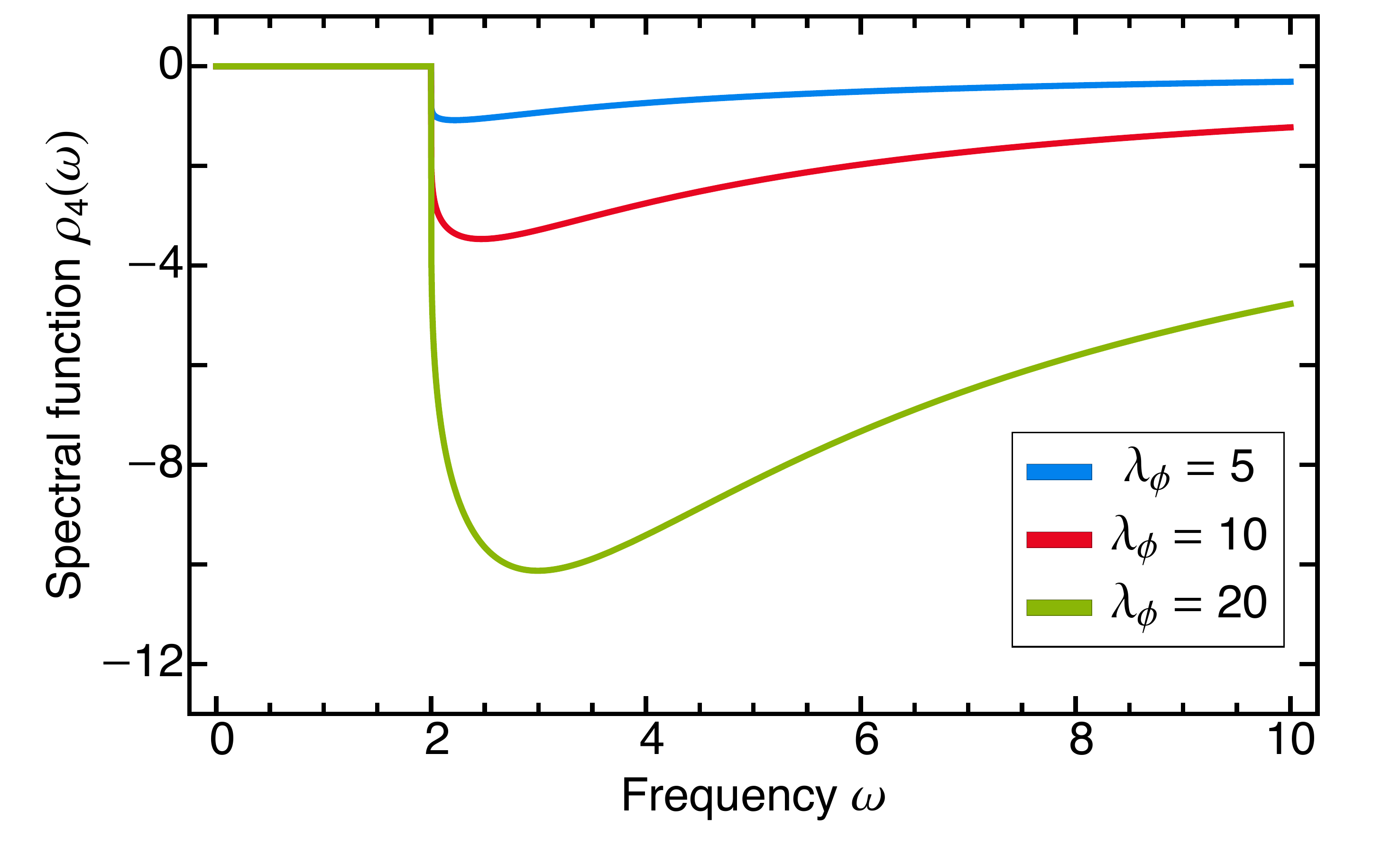}
	\caption{Left: Comparison of the momentum dependent (Euclidean) four-vertex $\Gamma^{(4)}(p)$ in its initial form~\labelcref{eq:bubble_resummation} with its spectral representation~\labelcref{eq:spec_rep_vert} for the coupling choices $\lambda_{\phi}$ = \{10,15,20\} using a classical propagator, i.e. a delta pole spectral function peaked at $m_{\text{pole}} = 1$. All dimensionful quantities were rescaled in units of the respective mass pole result. Right: Corresponding spectral functions of the four-point functions. The different height of the delta peaks encodes the magnitude of the residue relative to the other spectral functions. Also here, all dimensionful quantities were rescaled in units of the respective mass pole result.}
	\label{fig:four_vertex_spec_rep}
\end{figure*}
In our approximation with $\Gamma^{(3)}_{\text{pol}}(p) = \phi_0  \Gamma^{(4)}(p)$ with external momentum $p$, the vertices in the polarisation diagram are, as in the tadpole, proportional to the $s$-channel four-point vertex. For the sake of simplicity we also use  $\Gamma^{(3)}_{\text{pol}} = \phi_0  \Gamma^{(4)}(\omega=0)$. In any case, the spectral integrands $I_{\text{pol}}$ and $I_{\mathrm{sun}}$ of polarisation and sunset diagram remain the same as in \Cref{subsec:results_bare_vertices}. However, their prefactors $g_{\text{pol}}$ and $g_{\mathrm{sun}}$ are modified by the skeleton expansion, \cf \Cref{tab:prefactors_DSE}. 

As outlined, our solution method requires an analytic solution of the momentum integrals for all diagrams. In the current approximation this holds true for the polarisation and sunset diagram. It is not the case for the tadpole diagram because the loop-momentum is probing the non-trivial momentum structure of the resummed four-point function~\labelcref{eq:bubble_resummation}. This problem can be resolved if we can use a spectral representation for the resummed four-point function, which is discussed in the following section.

\subsection{Spectral representation for the four-point function} \label{subsubsec:spec_rep_vert}
\begin{figure*}[t]
	\centering
	\includegraphics[width=.48\linewidth]{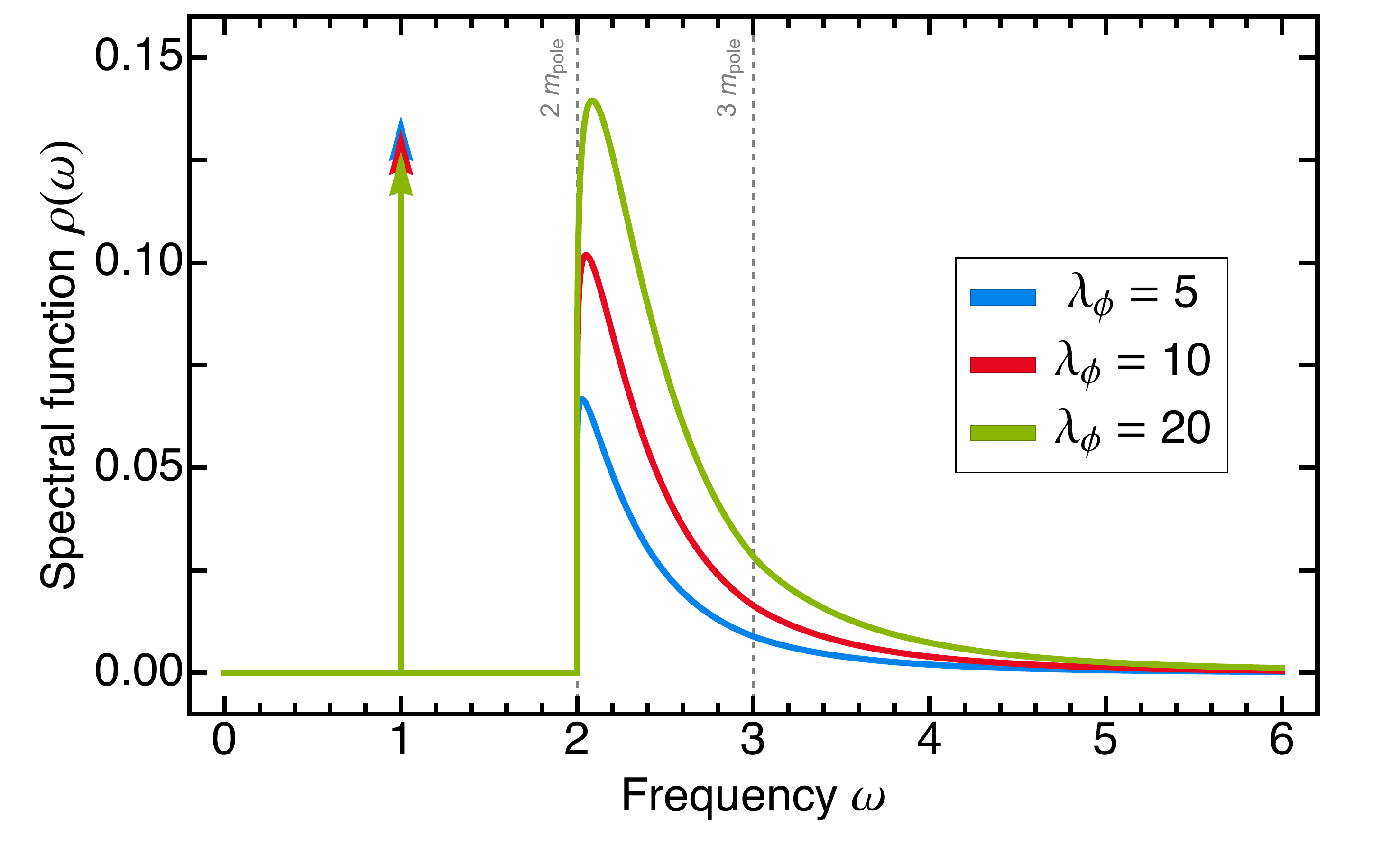}
	\includegraphics[width=.48\linewidth]{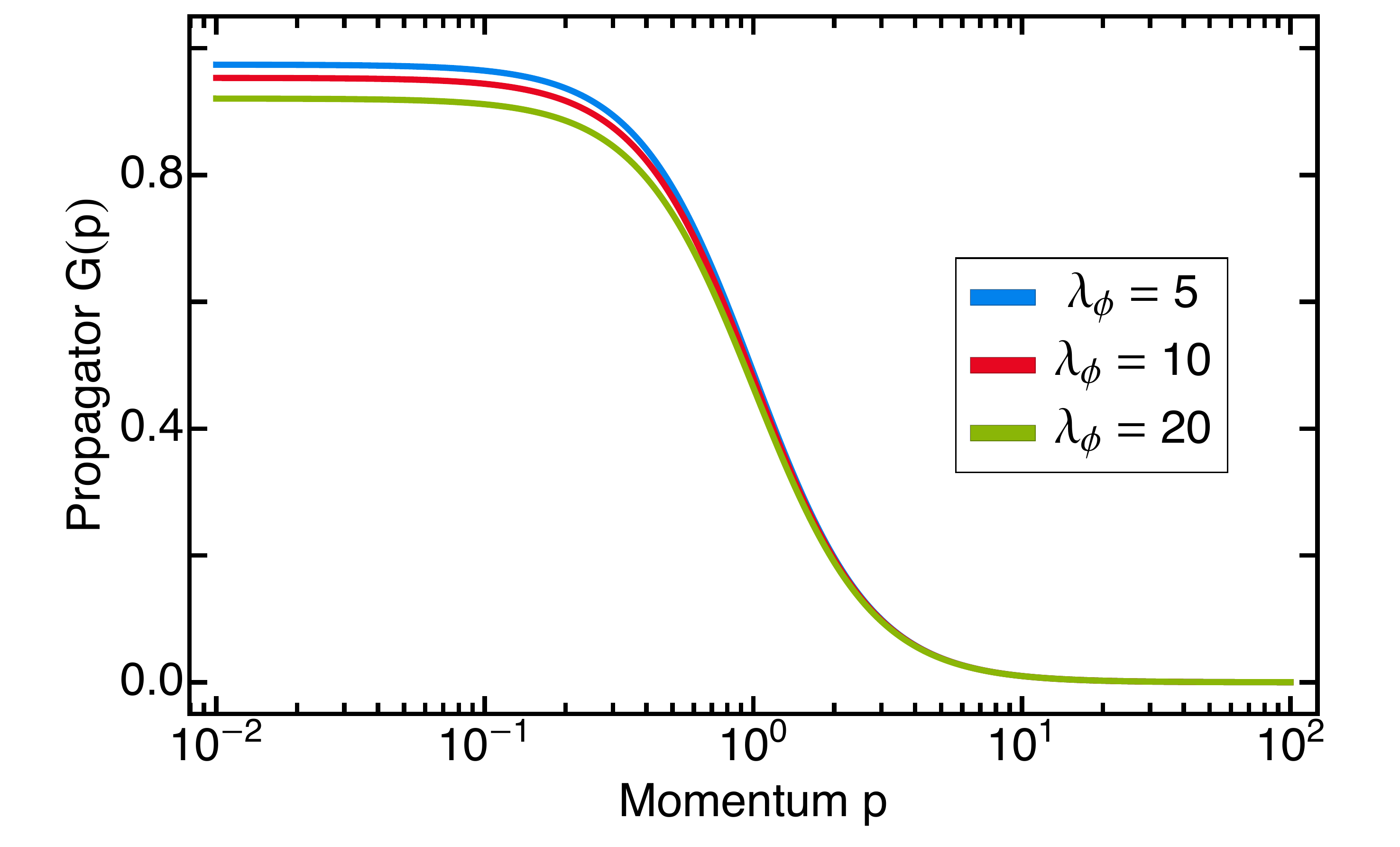}
	\includegraphics[width=.48\linewidth]{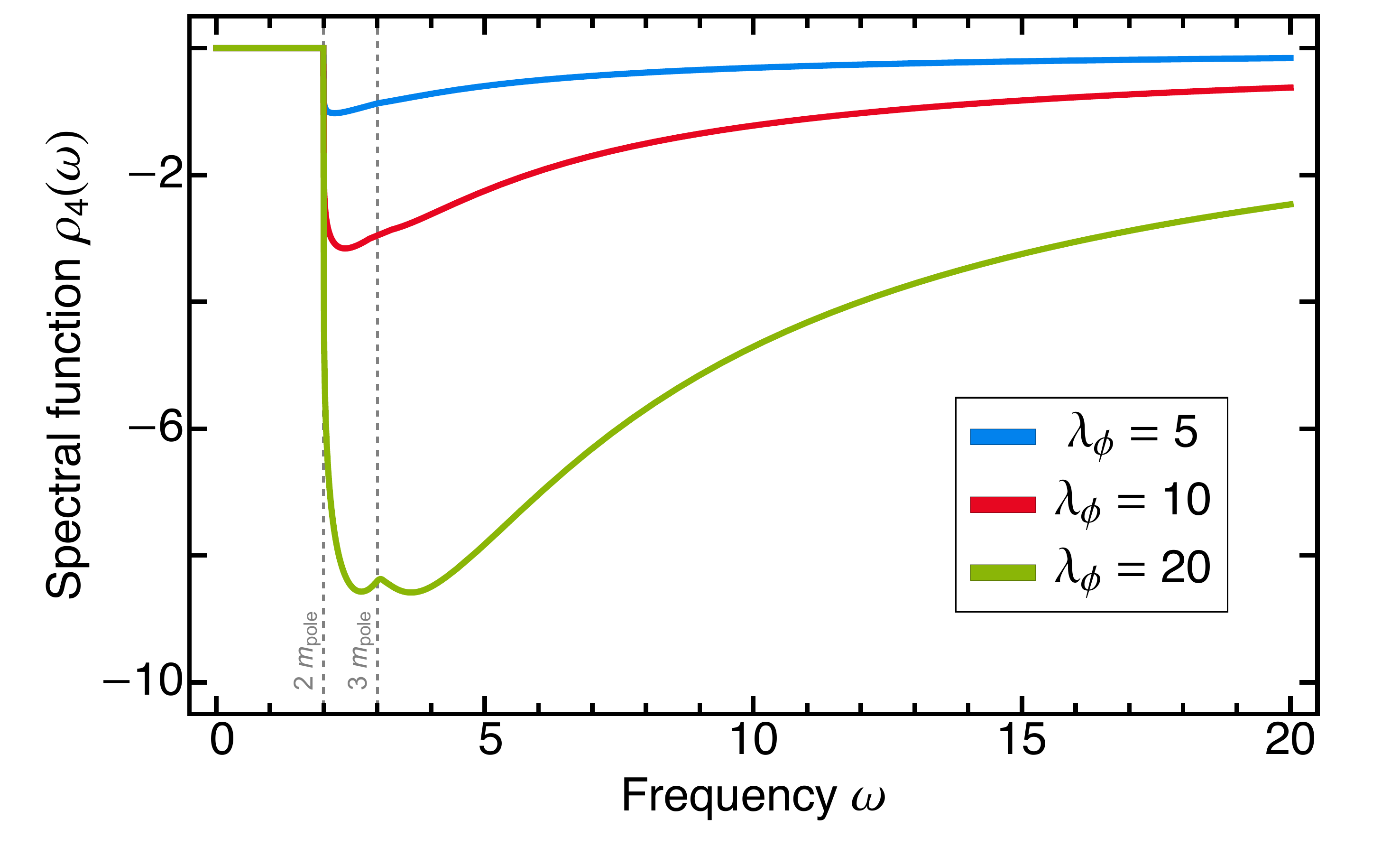}
	\includegraphics[width=.48\linewidth]{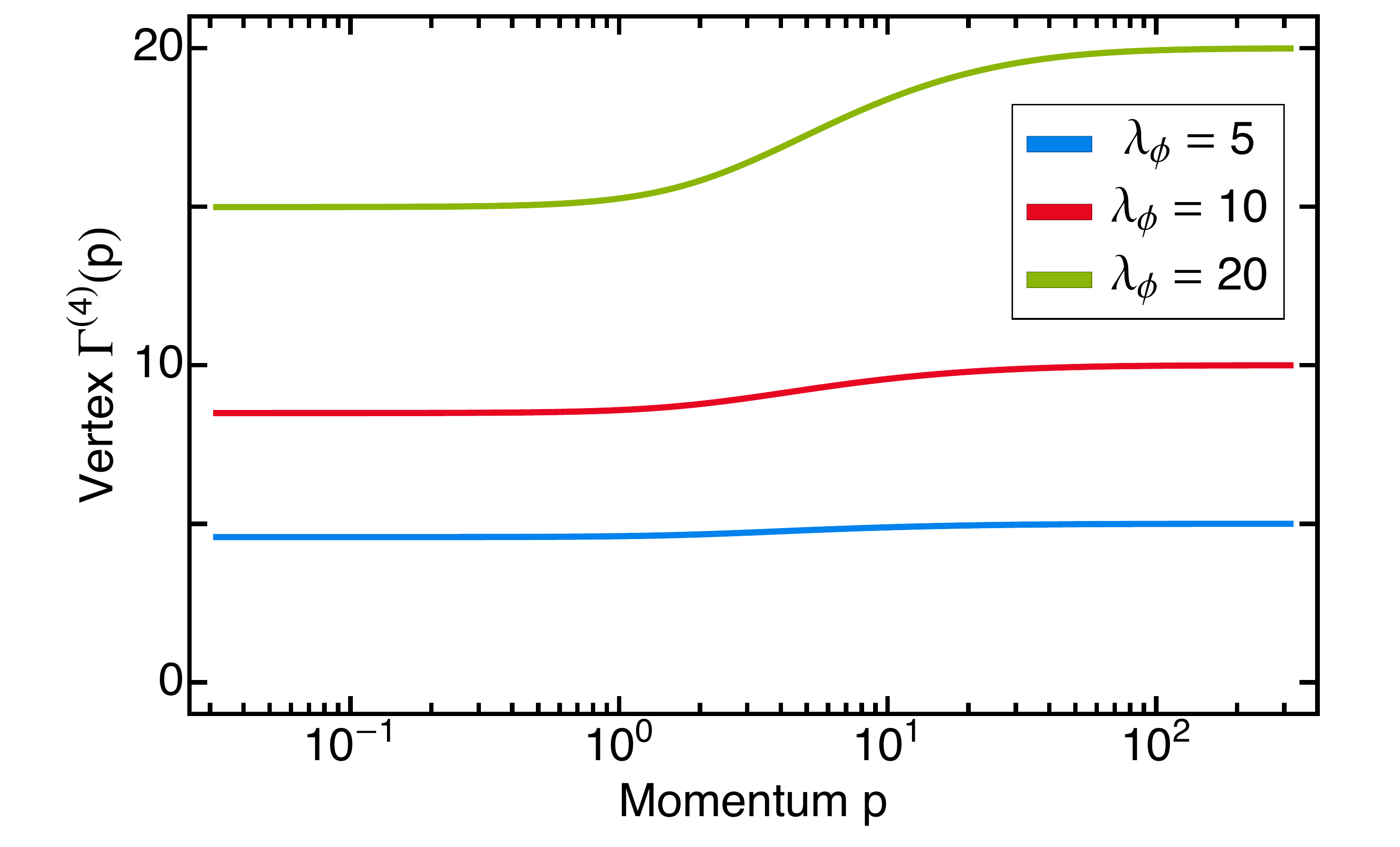}
	\caption{Results from the skeleton expanded DSE (comp.~\Cref{fig:DSE_2pt_full}) with a bubble resummed $s$-channel expansion of the four-point function for coupling choices ${\lambda}_\phi = 5, 10, 20$ using on-shell renormalisation~\labelcref{eq:RGOnshell}. The curves were rescaled by the respective mass poles. All vertices except for the tadpole one were approximated at $\omega=0$. TOP: Spectral function (left) and propagator (right). The weight of the continuous tail increases with coupling, the mass pole residue decreases. Higher $n$-particle onsets are not visible in the spectral function. The different height of the delta peaks encodes the magnitude of the residue relative to the other spectral functions. The propagators were computed by the K\"all\'{e}n-Lehmann spectral representation. Increasing coupling makes the propagators deviate more from the classical propagator, which approaches 1 at momentum $p \to 0$. BOTTOM: Four-vertex spectral function (left) and four-vertex (right).  Weight of the vertex spectral functions continuous tail increases with coupling. For the largest coupling choice, the three-particle onset is visible in the spectral function. The four-vertices were computed by their spectral-like representation (comp.~\labelcref{eq:spec_rep_vert}). The quantum corrections in the IR increase with coupling. All vertices approach their classical value in the UV.}
	\label{fig:results_skeleton}
\end{figure*}

While the existence and practical form of spectral representations for full four-point functions pose an intricate problem, spectral representations for (approximations of) single exchange channels of the four-point function can be derived. From a practical perspective we may treat such a channel similarly  to a propagator. This is well-motivated by considering that the resonant channels of a four-point function  correspond to particle exchange interactions. Technically, this can be made explicit by means of an Hubbard-Stratonovich transformation. In analogy to a propagator we can make the same ansatz for a spectral representation for a single channel of the resummed vertex 
\begin{equation} \label{eq:spec_rep_vert}
\Gamma^{(4)}(p) \; = \; \lambda_\phi \; + \int_{\lambda} \; \frac{\lambda\;\rho_4(\lambda)}{p^2+\lambda^2} \,,
\end{equation}
with
\begin{equation} \label{eq:spec_func_vert}
\rho_4(\omega) \; = \;  2 \; \mathrm{Im} \; \Gamma^{(4)}\big(-\mathrm{i} (\omega+\mathrm{i} 0^+) \big) \,.
\end{equation}
In \labelcref{eq:spec_rep_vert} the constant classical part $\lambda_\phi$ has to be separated. It has no spectral representation and does not need one for the present purpose. Indeed, classical vertices have been already considered in \Cref{subsec:results_bare_vertices}. 

Our results confirm that the analytic structure of the resummed vertex is compatible with \labelcref{eq:spec_rep_vert} and works well. This can be seen in \Cref{fig:four_vertex_spec_rep}, the computational details can be found in the next section, \Cref{subsec:solution_coupled_system}. The spectral function $\rho_4$ displayed in the right panel exhibits a continuous tail corresponding to the $\phi \phi \to \phi \phi$ scattering continuum for $\omega \geq 2m_{\text{pole}}$. The spectral representation~\labelcref{eq:spec_rep_vert} of the four-point function is used in the tadpole diagram in complete analogy to that of propagators. 

Importantly, the spectral representation of the vertex effectively just leads to another classical propagator with spectral mass $\lambda$ to the loop momentum integral. The momentum flowing through the four-vertex is the sum of the loop and external momentum. Thus, the momentum integral of the tadpole is identical to that of the polarisation diagram, since the internal line of the four-vertex carries $p+q$ and the initial internal line just the loop momentum $q$. The tadpole diagram can therefore be expressed as
\begin{equation} \label{eq:tadpole}
	D_{\mathrm{tad}}(\omega) = g_{\mathrm{tad}} \int_{\lambda_1,\lambda_2} \lambda_1 \lambda_2 \rho(\lambda_1) \rho_4(\lambda_2) I_{\mathrm{pol}}(\omega;\lambda_1,\lambda_2) \,.
\end{equation}
The spectral integral is logarithmically divergent, since the vertex spectral function $\rho_4$ drops off in the UV as $\lambda^{-1}$ (as opposed to $\rho \sim \lambda^{-2}$ in the UV). Again, we employ spectral BPHZ-renormalisation to subtract the zeroth order term of the Taylor expansion of $I_{\text{pol}}$. Finally, the renormalised diagram reads
\begin{align} \label{eq:tadpole_renormalized} \nonumber
D_{\mathrm{tad}}^{\mathrm{ren}}(\omega) = & \; g_{\mathrm{tad}} \int_{\lambda_1,\lambda_2}  \lambda_1 \lambda_2 \rho(\lambda_1) \rho_4(\lambda_2)  \\[1ex] 
 & \; \times \big[ I_{\mathrm{pol}}(\omega;\lambda_1,\lambda_2) - I_{\mathrm{pol}}(\mu;\lambda_1,\lambda_2) \big] \,.
\end{align}
%
\subsection{Results for the coupled system of propagator and vertices} \label{subsec:solution_coupled_system}
The DSE in the skeleton expansion is solved for the couplings also used in \Cref{subsec:results_bare_vertices}, ${\lambda}_\phi = 5, 10, 20$, measured in the pole mass $m_\text{pole} =1$. 

For the first iteration, initial choices $\rho_0, \rho_{4,0}$ for the spectral function of the propagator and that the four-point  vertex are required. For $\rho_0$ we use the classical spectral function \labelcref{eq:initial_guess_spec_func}, as already done in \Cref{subsec:results_bare_vertices}. For the spectral function of the four-point function $\rho_{4,0}$ compute it from the resummed representation of the four-point function, ~\labelcref{eq:bubble_resummation} with the initial choice of the spectral function of the propagator, $\rho_0$. This results in  
\begin{equation} \label{eq:initial_guess_vertex_spec_func}
	\rho_{4,0}(\omega) = 2 \; \text{Im} \; \frac{\lambda_\phi}{1 + \lambda_\phi \Pi_{\textrm{fish},0}(\omega)}  \,,
\end{equation}
with
\begin{align} \nonumber 
		\Pi_{\text{fish},0} = & \; \frac{1}{2}\int_{\lambda_1,\lambda_2} \lambda_1 \lambda_2 \rho_0(\lambda_1) \rho_0(\lambda_2) I_{\text{pol}}(\omega;\lambda_1,\lambda_2)  \\[1ex]
		= & \; \frac{1}{2} I_{\text{pol}}(\omega;m_{\text{pole}},m_{\text{pole}}) \,, 
\label{eq:initial_guess_self_energy}\end{align}
with $m_{\text{pole}} = 1$. While one could also simply take the classical vertex, the convergence speed and potentially also the convergence radius (coupling range) is increased by the improved choice \labelcref{eq:initial_guess_vertex_spec_func}. For further couplings $\lambda_\phi$ we take as initial choices $\rho_0$ and $\rho_{4,0}$ the full solutions $\rho$ and $\rho_{4}$ of the closest coupling value already computed. This procedure has already been used in  \Cref{subsec:results_bare_vertices}, and speeds up the convergence.  

The spectral function and propagator obtained from the coupled system of resummed four-point function and DSE are displayed in the top panels of~\Cref{fig:results_skeleton}. As for the case of bare vertices, shown in \Cref{subsec:results_bare_vertices}, we find a distinct one-particle mass pole as well as a scattering tail. The $\phi \to \phi \phi \phi$ onset is not visible in the spectral function for any of the coupling configurations. It can be seen that for increasing coupling $\lambda_\phi$, the tail of the spectral function becomes more enhanced, since the higher scattering states are more accessible. In turn, the mass pole residue decreases. The corresponding propagators in the right panel show similar behaviour as the for the DSE with bare vertices. The larger the coupling gets, the further the propagators deviate from the classical behaviour $G(p) \to 1$ for $p \to 0$.

In the bottom panels of~\Cref{fig:results_skeleton} we display the spectral function of the $s$-channel four-point function and the Euclidean four-point function itself. The consistency of the spectral representation has been discussed already in the previous section, and is confirmed numerically, see \Cref{fig:four_vertex_spec_rep}. Technically, the negativity of the spectral function displayed in \Cref{fig:results_skeleton} can be understood from the dominant quantum correction to the classical vertex, which is negative. On the conceptual side, within the Hubbard-Stratonovich transformation $\rho_4$ is related to minus the spectral function of the exchange particle. 

We find a continuous $2 \to 2$ scattering tail, starting at $2m_{\text{pole}}$ for all coupling choices. The spectral function is strongly enhanced with increasing coupling. Additionally, for larger coupling the spectral functions also clearly show the $1 \to 3$ scattering onsets starting at $3 m_\mathrm{pole}$, which was not visible in the propagator spectral functions (cf. top left panel of~\Cref{fig:results_skeleton}). By simple dimensional analysis it becomes clear that the higher $n$-particle thresholds in the propagator spectral function are suppressed by their respective energy threshold squared. This is not the case for the vertex spectral function: It decays with $\lambda^{-1}$, making the higher onsets less suppressed. In turn, the invisibility of four, five, and higher particle onsets is due to their decreasing amplitude, as every next higher onset comes with one additional loop. Further, we note that the visible size of the $1 \to 3$ scattering onset has its sole  origin in the tadpole diagram. This diagram contributes to the $1 \to 3$ scattering process due to the $s$-channel resummed four-point function, cf.~\Cref{subsubsec:tadpole_contribution_sunset}. The  contribution of the sunset itself is very suppressed in comparison. This points towards a general feature of our approximation: In the large coupling (massless) limit, the tadpole becomes the dominating diagram in the spectral function (for $\omega > 3 m_\mathrm{pole}$). We elaborate more on the massless limit in~\Cref{app:massless_limit}.  

While presenting the spectral and correlation functions in units of the fixed mass pole  $m_\mathrm{pole} = 1$ allows for a comparison of the relative strength of the different contributions, the approach to the massless limit is better studied if the results are presented in units of a uniform interaction strength. This is achieved by measuring all results in units of the coupling $\lambda_\phi$, see \Cref{app:massless_limit}. For the parameters $\lambda_\phi/m_\textrm{pole}=5,10, 20$ studied here this entails that we consider theories with the coupling $\lambda_\phi=1$ with pole masses $m_\textrm{pole} = 1/5, 1/10, 1/20$. Evidently, within these units the spectral functions pole position moves towards zero and the onset of the scattering states gets more pronounced, see \Cref{fig:results_skeleton_c}. 

\subsection{Self-consistent skeleton expansion} \label{subsec:self_consistent_polarisation}
\begin{figure*}[t]
	\centering		
	\includegraphics[width=.48\linewidth]{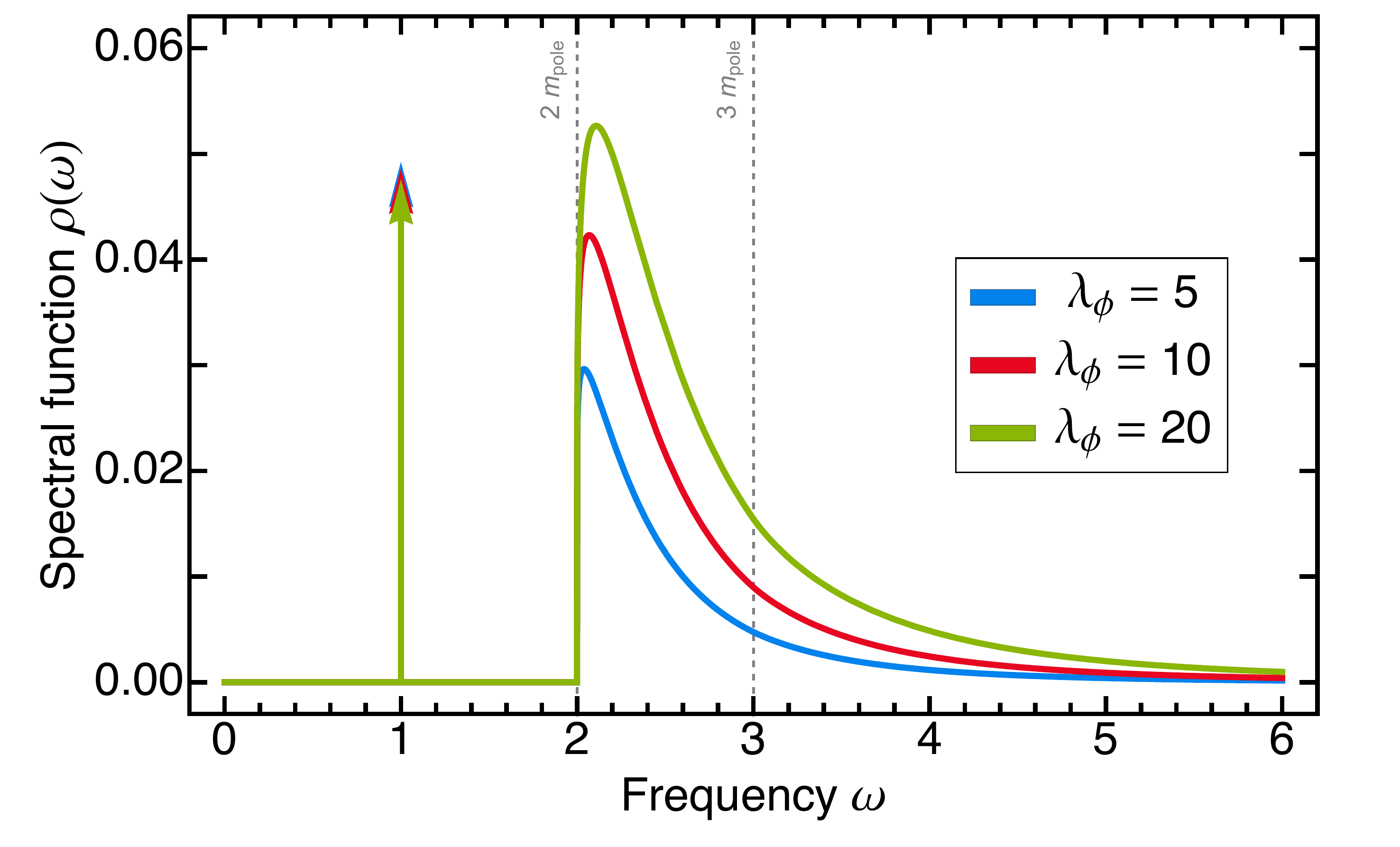}
	\includegraphics[width=.48\linewidth]{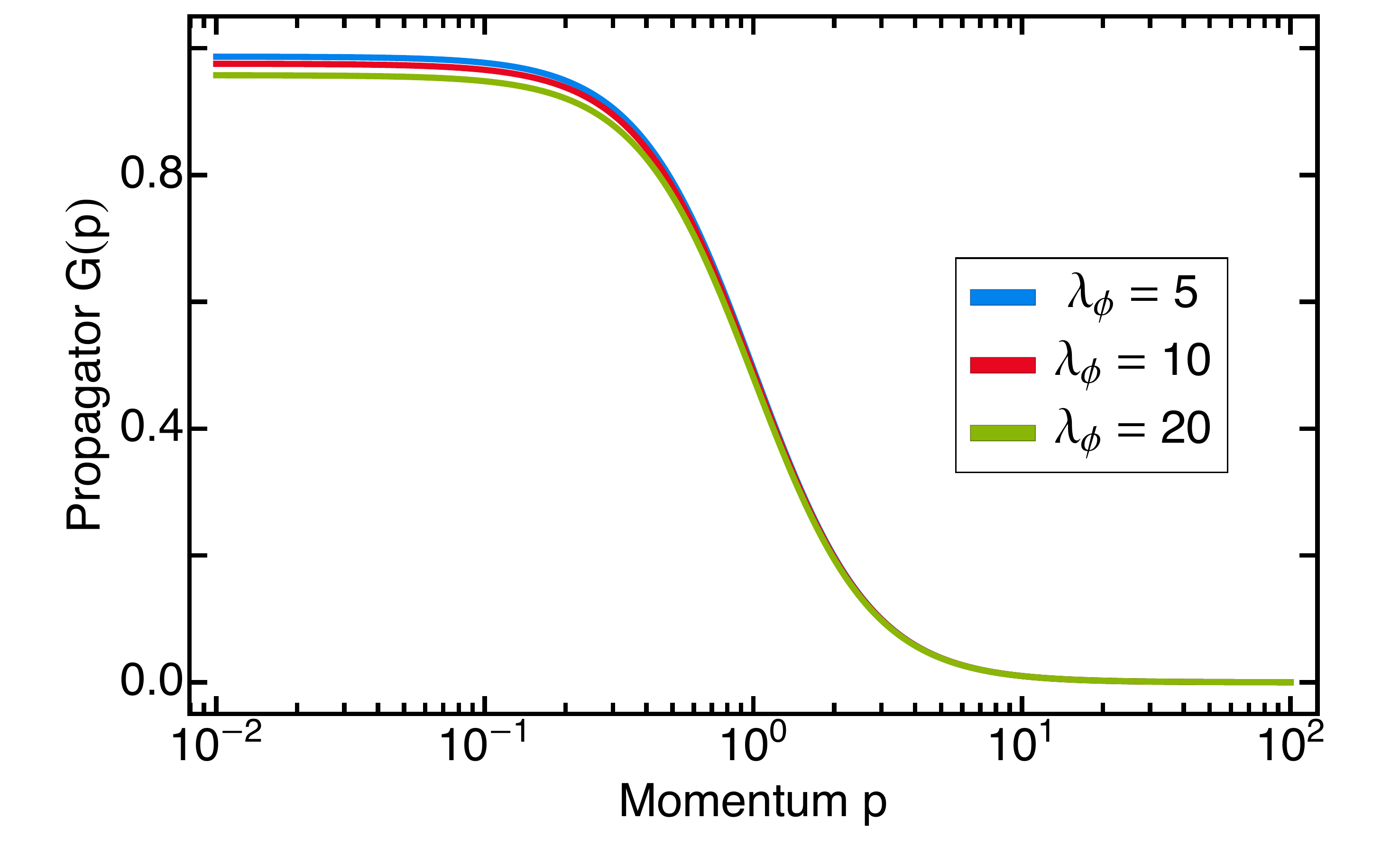}
	\includegraphics[width=.48\linewidth]{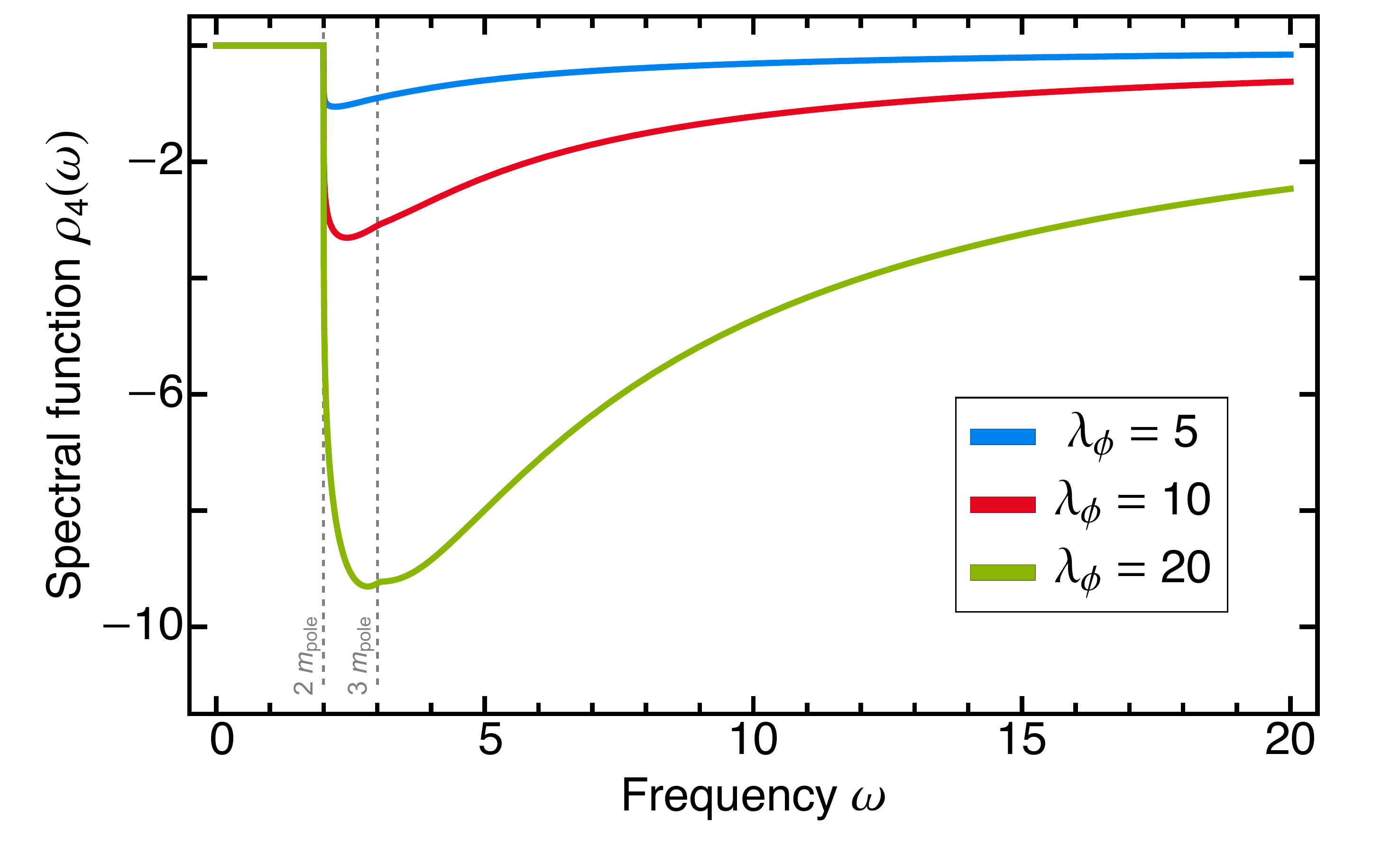}
	\includegraphics[width=.48\linewidth]{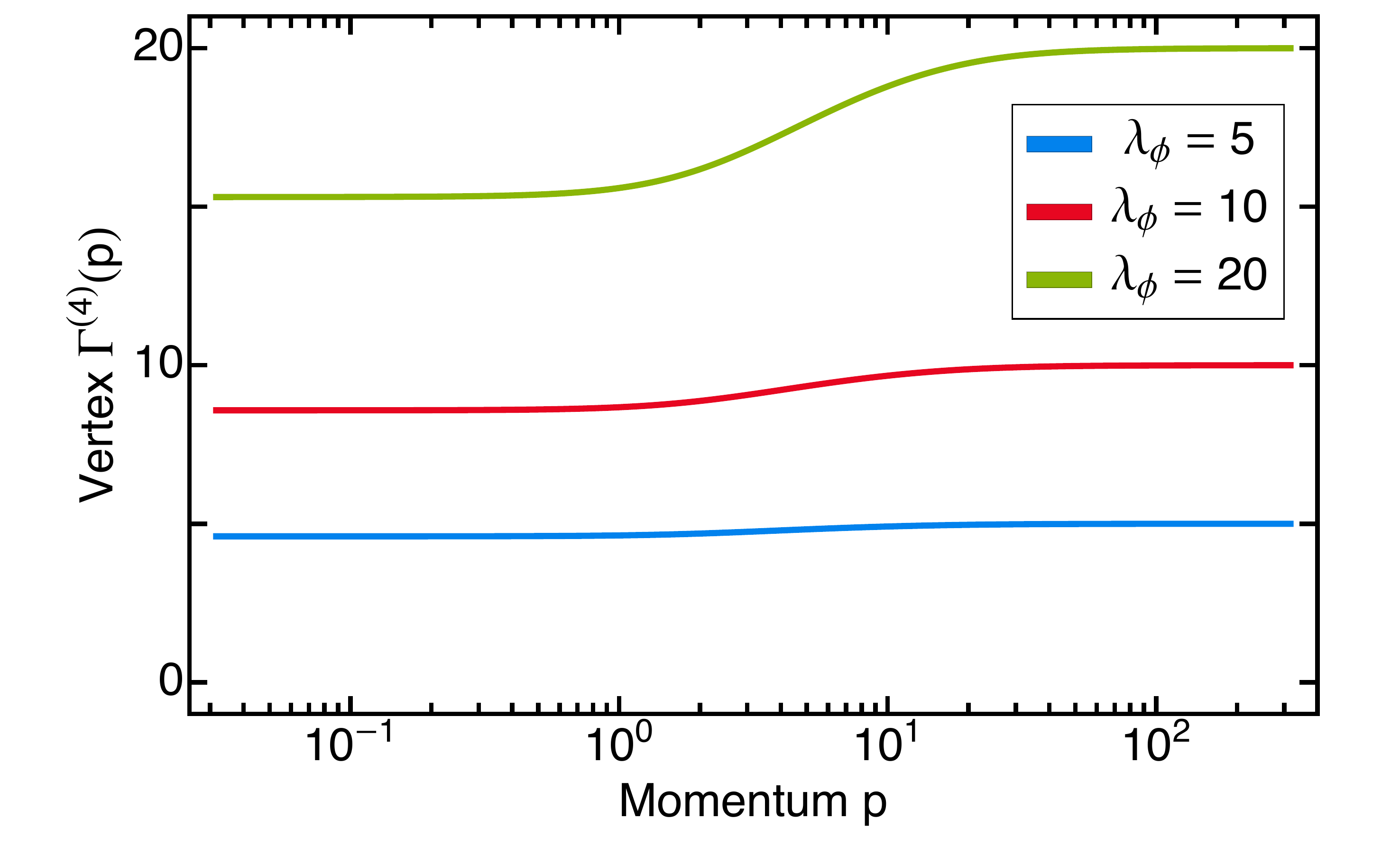}
	\caption{Results from the self-consistent skeleton expanded DSE (comp.~\Cref{fig:DSE_2pt_full}) with a bubble resummed $s$-channel expansion of the four-point function for coupling choices ${\lambda}_\phi = 5, 10, 20$ using on-shell renormalisation~\labelcref{eq:RGOnshell}. The polarisation diagram is here expressed through the $s$-channel four-vertex, cf. \labelcref{eq:bestofDSE}. The curves were rescaled by the respective mass poles. All vertices except for the tadpole one were approximated at $\omega=0$. TOP: Spectral function (left) and propagator (right). The weight of the continuous tail increases with coupling, the mass pole residue decreases. Higher $n$-particle onsets are not visible. The different height of the delta peaks encodes the magnitude of the residue relative to the other spectral functions. The propagators were computed by the K\"all\'{e}n-Lehmann spectral representation. Increasing coupling makes the propagators deviate more from the classical propagator, which approaches 1 at momentum $p \to 0$. BOTTOM: Four-vertex spectral function (left) and four-vertex (right).  Weight of the vertex spectral functions continuous tail increases with coupling. For the largest coupling choice, the three-particle onset is visible in the spectral function. The four-vertices were computed by their spectral-like representation (comp.~\labelcref{eq:spec_rep_vert}). The quantum corrections in the IR increase with coupling. All vertices approach their classical value in the UV.}
	\label{fig:results_skeleton_ex}
\end{figure*}

Within the current approximation, we have explicitly dropped the kite and double-bubble diagrams. Implicitly we have also dropped the squint diagram, that corresponds to vertex corrections to the three-point function in the vacuum polarisation: we have used two dressed three-point vertices that are derived from the bubble-resummation of the four-point function. Such an approximation of the three-point function does not contain the squint topology.

An alternative approximation of the Dyson-Schwinger equation is derived as follows: We start from the initial, full propagator DSE (\Cref{fig:DSE_2pt}) and consider it in a diagrammatic expansion in orders of the constant field $\phi_0$. In our approximation, these higher orders come via the three-vertices $\Gamma^{(3)}(p) = \phi_0 \Gamma^{(4)}(p)+O(\phi_0^3)$. Acting on the DSE with two derivatives w.r.t.\ the constant field $\phi_0$ and multiplying by $\phi_0^2/2$ afterwards, one finds the schematic relation
\begin{equation} \label{eq:pol_squint_gamma4}
O_{\Gamma^{(2)},\text{diag}} \big[ \phi_0^2 \big] (p) = \frac{1}{2} \phi_0^2 \big[ \Gamma^{(4)}(p,-p,0,0) - \lambda_\phi \big] \;,
\end{equation}
where $O_{\Gamma^{(2)},\text{diag}} \big[ \phi_0^2 \big]$ represents all diagrams in the propagator DSE with two external constant field legs, including their initial prefactors. This includes the squint diagram and the vacuum polarisation. 

On the right hand side of~\labelcref{eq:pol_squint_gamma4}, we have the full four-point function subtracted by its classical value. We also made explicit the specific momentum dependence of the four-point function. By differentiating twice w.r.t.\ to the momentum independent field $\phi_0$, $\Gamma^{(4)}$ only depends on one external momentum $p$.

This entails that we can re-express all diagrams of $O_{\Gamma^{(2)},\text{diag}} \big[ \phi_0^2 \big]$ through an $s$-channel four-point function. What is missing is the classical part of the vertex DSE. It is included by adding and subtracting the classical vertex contribution multiplied by an appropriate prefactor involving the constant field, $\tfrac{1}{2} \phi_0^2 \lambda_\phi$ to the propagator DSE. This leads us to \labelcref{eq:pol_squint_gamma4}. Evidently, the additional constant part $-\tfrac{1}{2} \phi_0^2 \lambda_\phi$ in \labelcref{eq:pol_squint_gamma4} is absorbed in the mass renormalisation. 

The expectation value $\phi_0^2$ can be expressed in terms of $\Gamma^{(2)}(0)=m_\textrm{cur}^2$ and $\Gamma^{(4)} [0]$. This leads us to
\begin{align} \label{eq:polGamma4fin}
\frac12 \phi_0^2 \,\Gamma^{(4)}(p)= \frac32 m^2_\textrm{cur}\frac{\Gamma^{(4)}(p)}{\Gamma^{(4)}(0)}\,, 
\end{align}
which has the scaling of a two-point function, and reduces to $3/2 m_\textrm{cur}^2$ at vanishing momentum. Making use of~\labelcref{eq:pol_squint_gamma4} and \labelcref{eq:polGamma4fin}, we are led to the DSE for the propagator with Minkowski frequencies $\omega$, 
\begin{align} \label{eq:bestofDSE} \nonumber 
\Gamma^{(2)}(\omega)= &\,-\omega^2 +m_\textrm{pole}^2 + D^\textrm{ren}_\text{tad}(\omega) + D^\textrm{ren}_\text{sun}(\omega) \\[1ex]
&\, +\frac32 m^2_\textrm{cur}\left[\frac{\Gamma^{(4)}(\omega)-\Gamma^{(4)}(m_\textrm{pole})}{\Gamma^{(4)}(0)} \right]\,, 
\end{align}
and $i=\textrm{tad,sun}$. 
Note that the polarisation and squint diagram have been absorbed into the last term of~\labelcref{eq:bestofDSE} proportional to $m_\mathrm{cur}^2$ as a result of~\labelcref{eq:pol_squint_gamma4}. Due to the on-shell renormalisation condition \labelcref{eq:RGOnshell}, all renormalised diagrams vanish at $\omega=m_\textrm{pole}$, that is $D_i^\textrm{ren}(m_\textrm{pole})=0$. In summary, \labelcref{eq:bestofDSE} is exact up to higher orders of $\phi_0^2$, leaving us with a self-consistent systematic expansion scheme. The self-consistency refers to the fact that in the present order in $\phi_0^2$, the polarisation diagram is given exactly in terms of the four-point vertex. Therefore, approximations to the latter are transported to the former. 

\subsubsection{Vertex-approximation in the self-consistent skeleton expansion} \label{subsubsec:vertex_approx_self_consistent}

As discussed above, within the self-consistent DSE in~\labelcref{eq:bestofDSE}, it suffices to specify the approximation for the four-point function. Here we again resort to the bubble-resummed four-point function of~\labelcref{eq:bubble_resummation}, already  used in the previous section. 
This approximation of the four-point function neglects in particular contributions in the DSE of the four-point function that originate in the squint diagram. The self-consistency of \labelcref{eq:bestofDSE} is reflected in the fact that the contribution of the polarisation diagram is given by its bubble-resummed  $1/2 \phi_0^2  \Gamma^{(4)}$, 
\begin{align} \label{eq:polGamma4}
\frac{1}{2} \phi_0^2 \,\Gamma^{(4)}(p) = \frac{1}{2} \phi_0^2\left[ \lambda_\phi - \lambda_\phi \Pi_\textrm{fish} \Gamma^{(4)}(p) \right] \,. 
\end{align}
As for the four-point function, \labelcref{eq:polGamma4}, lacks the contributions from the squint diagram. Note also that these contributions are related to the $u,t$-channel. Hence, with \labelcref{eq:polGamma4} we consistently neglect the squint topology in the DSE if assuming dominance of the $s$-channel vertex corrections for the four-point vertex. We emphasise that the assumption of $s$-channel dominance is well-supported in the large-$N$ limit, but less so in the present $N=1$ case.  

Diagrammatically, the DSE is still represented by~\Cref{fig:DSE_2pt_full}, with the polarisation diagram given by the last term in~\labelcref{eq:bestofDSE}. The prefactors of tadpole and sunset diagram are identical to that in the standard skeleton scheme used  in~\Cref{subsec:solution_coupled_system}. They are listed in~\Cref{tab:prefactors_DSE}. In the tadpole diagram, the four-point function again enters via its spectral representation~\labelcref{eq:spec_rep_vert}.

\subsubsection{Results} \label{subsubsec:results_self_consistent}

Numerical results for the self-consistent skeleton scheme are displayed in~\Cref{fig:results_skeleton_ex}. The propagator spectral function is similar in shape to that obtained in the standard skeleton approximation, see~\Cref{fig:results_skeleton}. However, the magnitude of the scattering tail close to the threshold is roughly a third for all coupling choices. Comparing the two skeleton schemes one realises that the correct momentum scaling behaviour of the standard skeleton expansion came at the price of two dressed vertices. In turn, the self-consistent skeleton expansion has the correct momentum scaling as well as the correct vertex strength. Arguably, this property is particularly important in the vicinity of $s$-channel resonances or for asymptotically large couplings. 

Higher $n$-particle onsets are not visible in the propagator spectral function similarly to the results in the standard skeleton expansion. We emphasise again, that they are present nevertheless, as well as easily accessible in the present spectral approach. The three-particle onset can again be seen in the vertex spectral function (bottom left panel of~\Cref{fig:results_skeleton_ex}), although it is less pronounced as in~\Cref{fig:results_skeleton}. The magnitude of the vertex spectral functions matches very well in the two different schemes however. 

The corresponding Euclidean correlation functions are shown in the top and bottom right panel. Both receive slightly less quantum corrections as for the plain skeleton expansion. For the propagator this is quite clear from the much smaller spectral functions in the self-consistent approximation. For the  four-point vertex, the differences are less pronounced and we refrain from discussing them. For a more detailed discussion of the general features, see~\Cref{subsec:solution_coupled_system}.

\subsubsection{Low-lying bound state close to phase transition} \label{subsubsec:bound_state}
Lattice calculations~\cite{Agostini:1996xy, Caselle:1999tm,Caselle:2001im} show an additional low-lying excitation in the spectrum of the scalar $\phi^4$-theory in $d=2+1$ close to the phase transition with a mass of $m \approx 1.8 m_\mathrm{pole}$. This state has been interpreted as a bound state of the fundamental excitation in \cite{Caselle:2001im}. It is also been observed within the recent functional RG study~\cite{Rose:2016wqz}. The approximation scheme underlying the Euclidean computation done in \cite{Rose:2016wqz}. is close in spirit to the $s$-channel approximation scheme used in the present work in Minkowski space-time In \cite{Rose:2016wqz}, the spectral function of the propagator, numerically reconstructed from Euclidean data, shows a bound state close to the phase transition at a mass ratio consistent with that found in the lattice studies. However, further away from the phase transition, the clear signal of the bound state is lost in~\cite{Rose:2016wqz}. 

The present direct computation of spectral functions is not performed close to the phase transition, that is for $m_\textrm{pole}\to 0$. This regime will be studied elsewhere within the self-consistent skeleton expansion developed in the present section. Our results for $m_\textrm{pole}\neq 0$ indicate that the bound state may indeed only exist close to the phase transition. However, the present $s$-channel approximation does not allow for a fully conclusive statement, as the latter requires a multi-channel analysis. Yet, the $s$-channel resummation typically captures the dominant  resonances and is trustworthy as next-to-leading order in a $1/N$-expansion in the large-$N$ limit. The spectral properties of $O(N)$ models as well as the phase transition regime will be studied elsewhere.

\section{Conclusion} \label{sec:Conclusion}

In this work, we developed a \textit{spectral} functional approach for the direct non-perturbative computation of real-time (time-like) correlation functions. This approach is based on a novel renormalisation scheme called \textit{spectral renormalisation}: This renormalisation scheme is based on the use of spectral representations and dimensional regularisation. The spectral representation allows to perform the momentum integrals within dimensional regularisation. This leaves us with the spectral integrals, whose spectral divergences can be renormalised within dimensional regularisation as well, see \Cref{subsec:spectral_renormalisation}. This scheme is called \textit{spectral dimensional renormalisation}, and respects all symmetries of the theory at hand that are maintained within dimensional regularisation. The latter set also includes gauge symmetries, and hence spectral dimensional renormalisation as developed here is a manifestly gauge-invariant renormalisation scheme. 

The renormalisation step can also be done with a standard subtraction procedure within a Taylor expansion in momenta as done in BPHZ-renormalisation This scheme is called \textit{spectral BPHZ-renormalisation}. It maintains less symmetries than spectral dimensional renormalisation and in particular requires counterterms in gauge theories that break gauge symmetry (or rather BRST-symmetry). The appeal of spectral BPHZ renormalisation lies in its relative numerical simplicity. 

In summary, \textit{spectral renormalisation} allows for direct access to the real momentum axis by analytic continuation as a result of the fully analytic solution of all momentum integrals, while maintaining all symmetries of the theory at hand. 

We performed explicit, non-perturbative computations within the spectral Dyson-Schwinger approach to the scalar $\phi^4$-theory in 2+1 space-time dimensions. Our results include the spectral function of the scalar propagator and that of the four-point function ($s$-channel), and are obtained by solving the spectral DSE recursively. We have first considered the approximation with classical vertices and full propagators, but including the two-loop diagrams in the DSE. The resulting spectral functions show a distinct one-particle pole and a clear scattering tail with onset at twice the pole mass. The spectral function contains all higher scattering thresholds, which are easily accessible due to the analytic nature of the momentum integrations. For increasing couplings, the $1 \to 3$ scattering onset becomes more pronounced. 

\begin{figure*}[t]
	\centering
	\includegraphics[width=.48\linewidth]{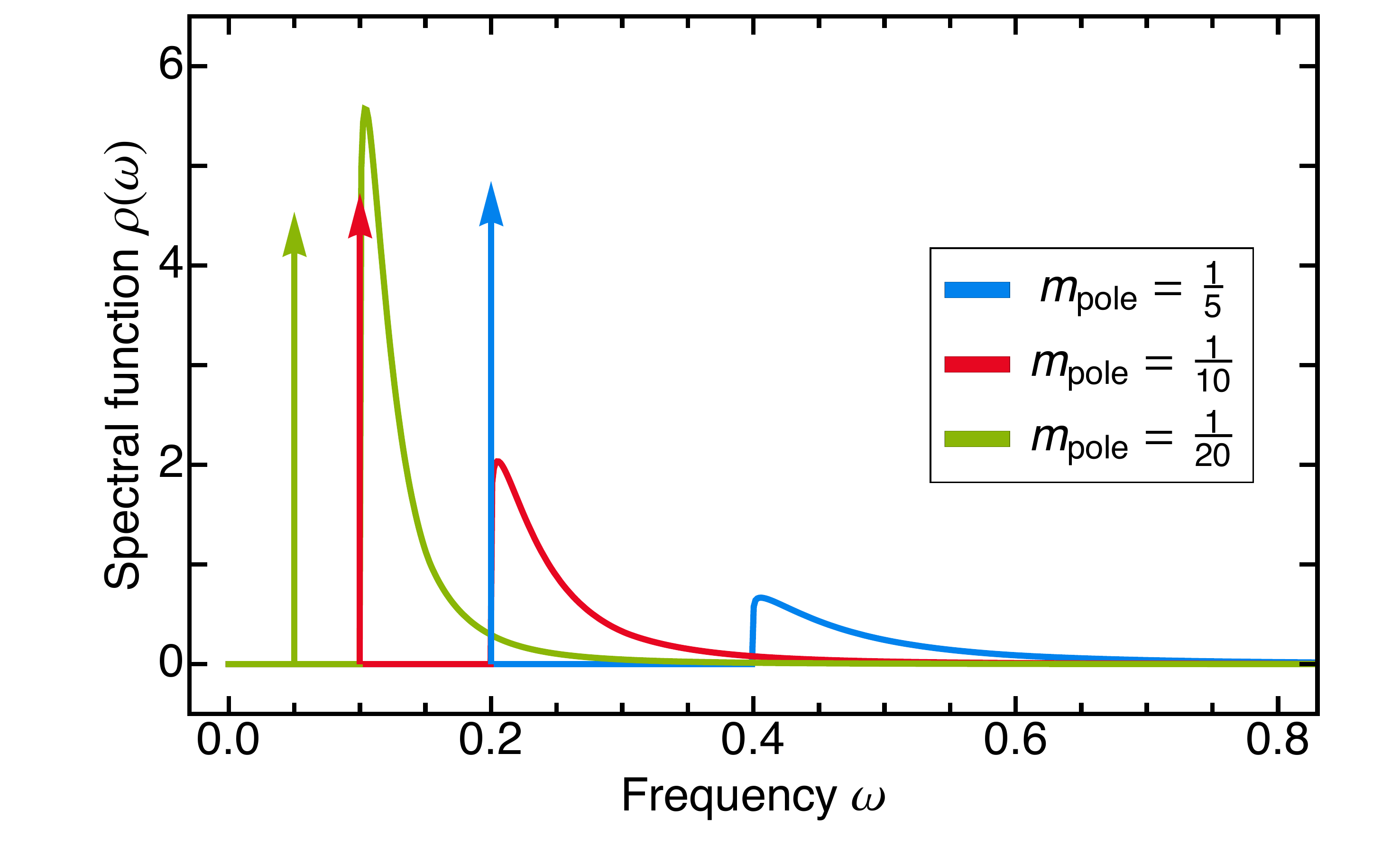}
	\includegraphics[width=.48\linewidth]{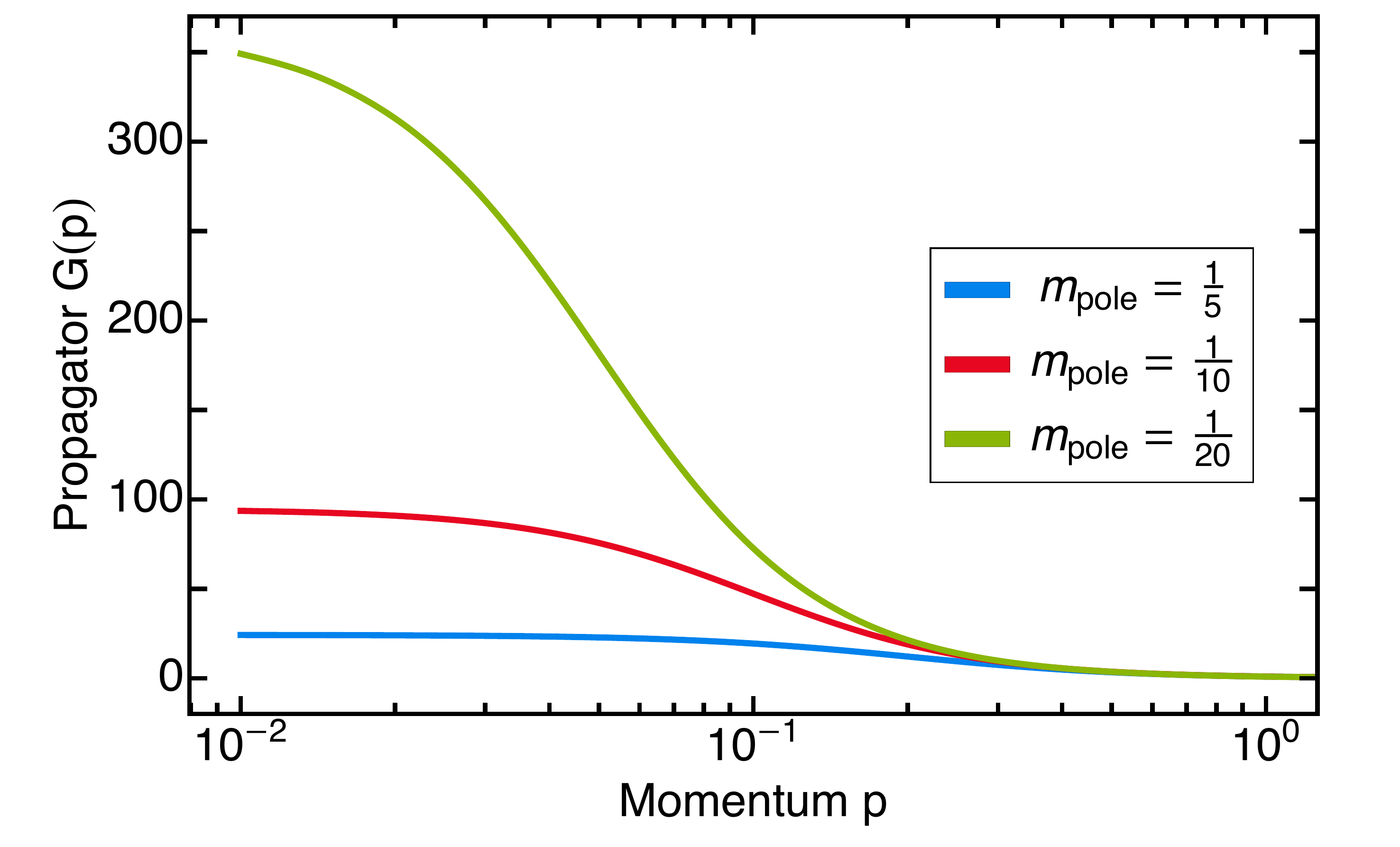}
	\includegraphics[width=.48\linewidth]{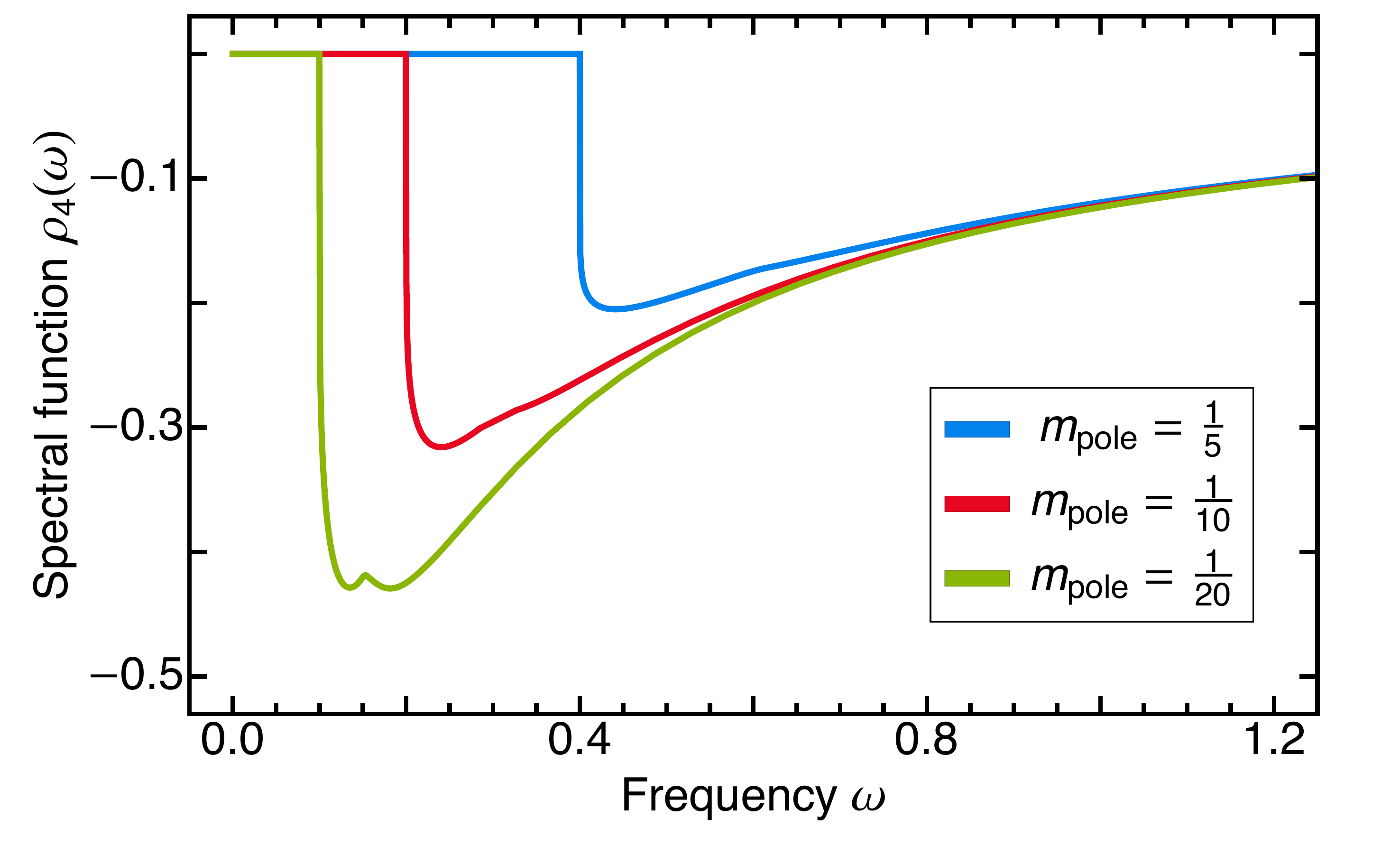}
	\includegraphics[width=.48\linewidth]{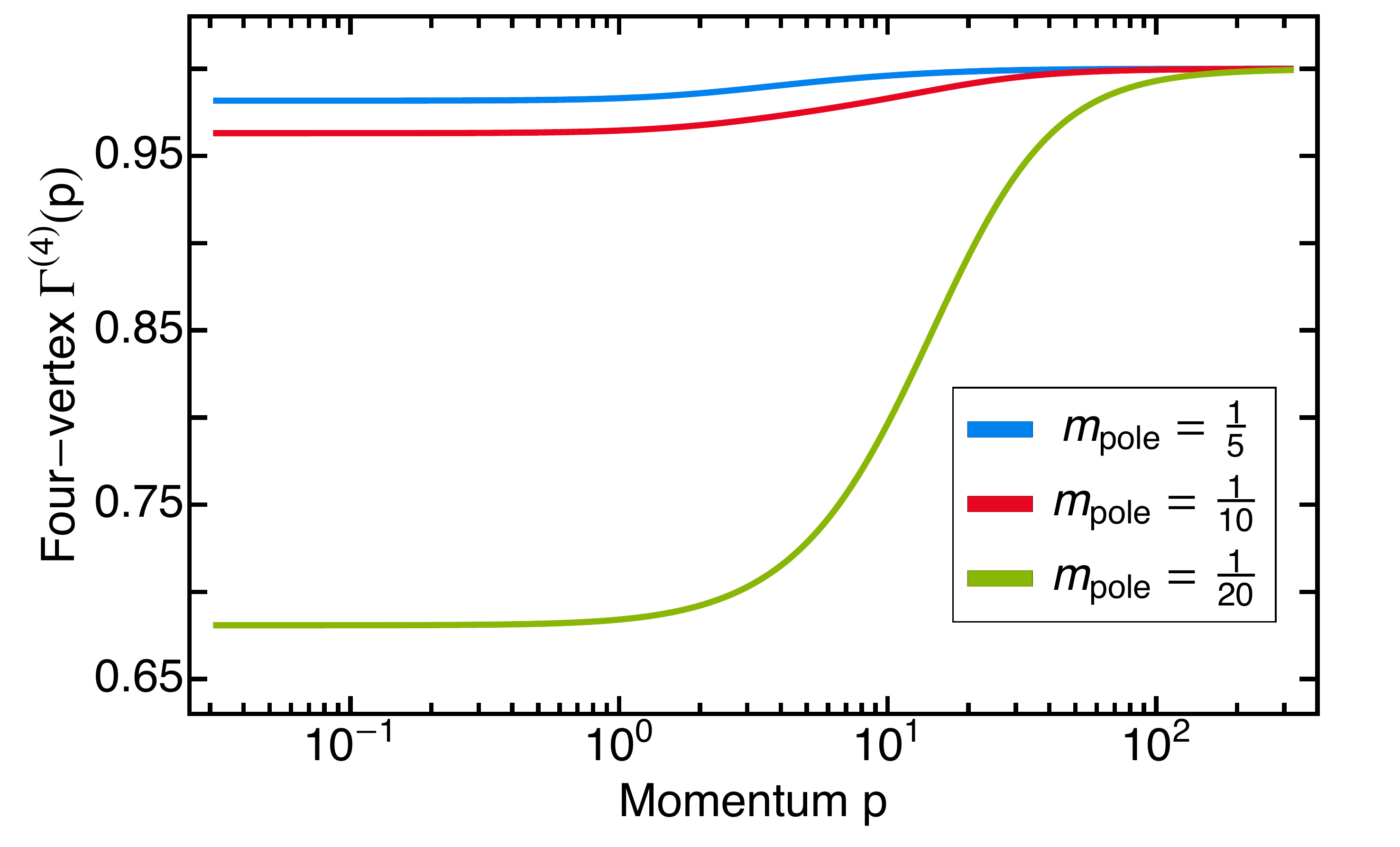}
	\caption{Results from the skeleton expanded DSE (comp.~\Cref{fig:DSE_2pt_full}) with a bubble resummed $s$-channel expansion of the four-point function for coupling choices ${\lambda}_\phi = 5, 10, 20$ using on-shell renormalisation~\labelcref{eq:RGOnshell}. All vertices except for the tadpole one were approximated at $\omega$ = 0. The curves were rescaled by their respective coupling parameters, i.e. $m_{\mathrm{pole},i} = \frac{1}{\lambda_{\phi,i}}$ and consequently $\lambda_\phi = \frac{\lambda_{\phi,i}}{\lambda_{\phi,i}} = 1$ for all curves. TOP: Spectral function (left) and propagator (right). The weight of the continuous tail decreases with larger mass pole, the mass pole residues increases. The different height of the delta peaks encodes the magnitude of the residue relative to the other spectral functions. The propagators were computed by the K\"allen-Lehmann spectral representation. The pole contributions are dominant, since for smaller mass pole the propagator is strongly enhanced.}
	\label{fig:results_skeleton_c}
\end{figure*}

We also considered an approximation with non-perturbative vertices, based on a skeleton expansion scheme. The respective four-point function was given by an $s$-channel bubble-resummation. The  propagator spectral function again shows a distinct one-particle pole and a continuous scattering tail. The spectral function of the resummed four-vertex features a scattering tail as well and additionally shows a distinct onset for the $1 \to 3$ scattering process. For larger couplings, the fully non-perturbative nature of the approximation leads to large, though only quantitative differences compared to the classical vertex computation. 

In the last part of this work,~\Cref{subsec:self_consistent_polarisation}, we have developed a self-consistent skeleton expansion scheme. The self-consistency was obtained by relating a class of diagrams with three-point functions to the $s$-channel four-point function. This entails that the approximation used in the computation of the four-point function is also used within the three-point function diagrams. The results are qualitatively similar to that of the standard skeleton scheme. The magnitude of the scattering tail of the propagator spectral function turns out much smaller in the upgraded scheme, and indicates an overestimation of the polarisation diagram before. We expect that the use of such a self-consistent scheme is important close to the phase transition of the theory or, more generally, in the presence of resonant $s$-channel interactions.  

In conclusion, we have developed and put to work a fully non-perturbative spectral functional approach to the computation of real-time correlation functions. The approach was successfully tested within the scalar $\phi^4$-theory in 2+1 dimensions. It is specifically attractive in gauge theories, as spectral dimensional renormalisation preserves gauge invariance. We hope to report on respective results soon. 

\section{Acknowledgments} \label{Acknowledgments}
\begin{figure*}[t]
	\centering
	\includegraphics[width=.48\linewidth]{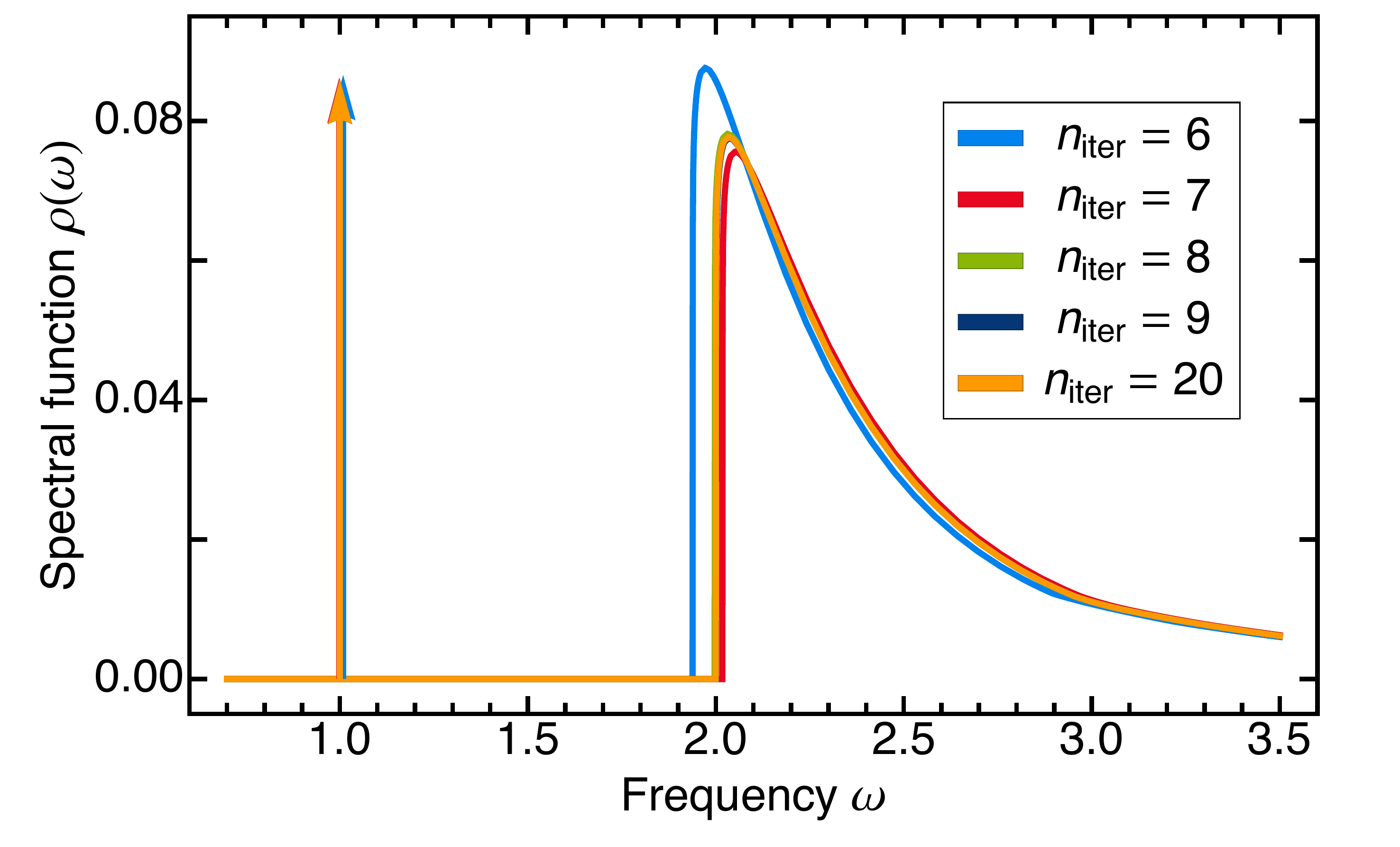}
	\includegraphics[width=.48\linewidth]{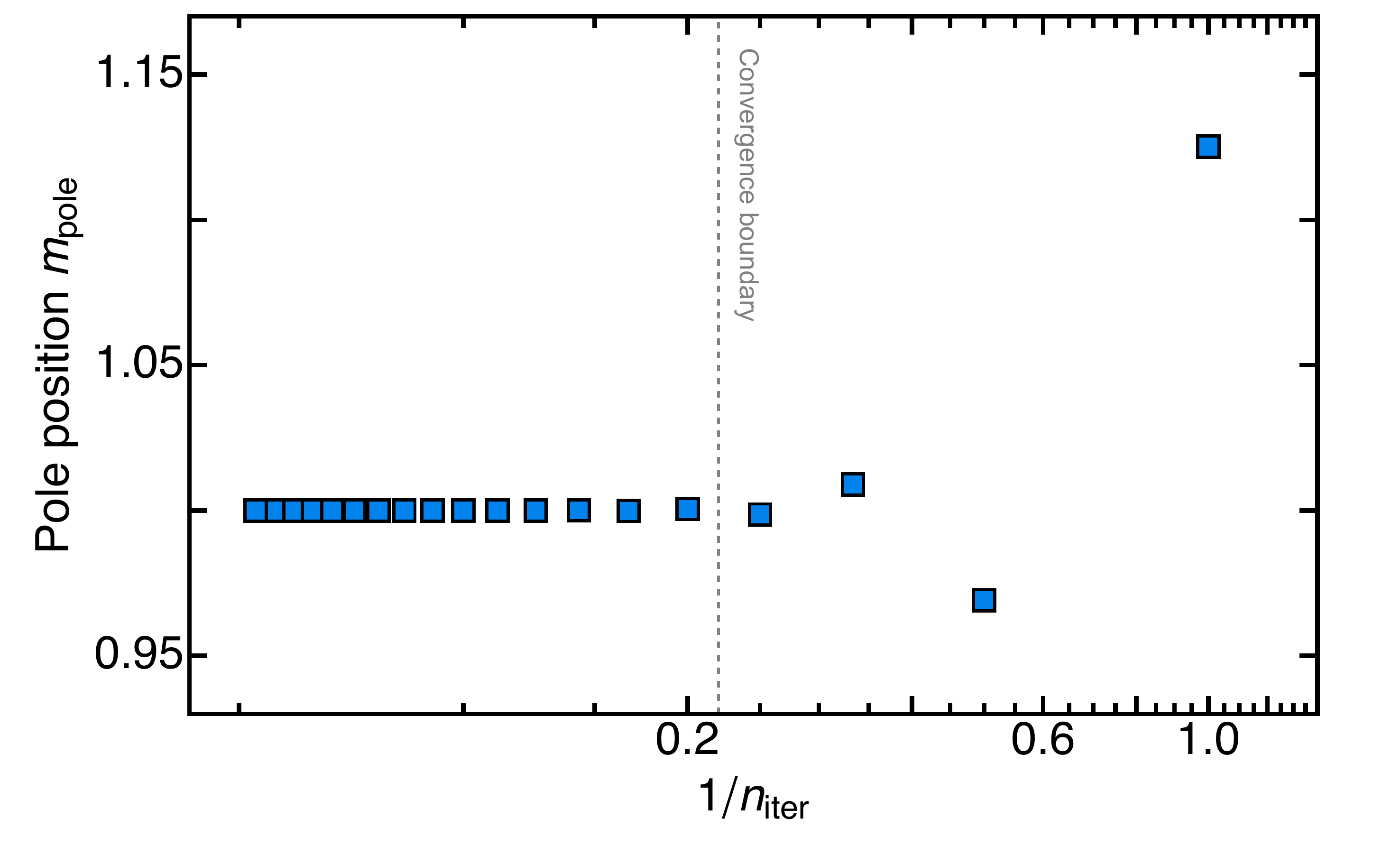}
	\caption{Example of a convergent iteration of the scalar DSE with classical vertices. The curves were \textit{not} rescaled by the respective mass poles to demonstrate convergence of the iteration also for the pole mass. Instead, renormalisation was done at $\mu = 0$. Units thus given by the input parameters. The spectral functions are alternating, approaching the final orange curve (left). 20 iterations have been performed. The curves from iterations 10 to 19 were left out as they were graphically indistinguishable from the final curve. An iterative behaviour as displayed is taken taken to be convergent, \ie signalling a solution to the DSE where left and right hand side of the equation coincide. The corresponding pole positions also converge very quickly (right). Left of the convergence boundary, assuming a (relative) precision of $10^{-4}$, all points are identical.}
	\label{fig:specFuncConvPlot}
\end{figure*}
We thank G.~Eichmann, C.~Fischer, M.~Hasenbusch, C.~Jung, P.~Lowdon, N.~Stamatescu, M.~Salmhofer and L.~von Smekal for discussions. This work is supported by the Deutsche Forschungsgemeinschaft (DFG, German Research Foundation) under Germany's Excellence Strategy EXC 2181/1 - 390900948 (the Heidelberg STRUCTURES Excellence Cluster) and under the Collaborative Research Centre SFB 1225 (ISOQUANT) and the BMBF grant 05P18VHFCA.


\bookmarksetup{startatroot}
\appendix

\section{Massless limit} \label{app:massless_limit}

The scalar $\phi^4$-theory in 2+1 dimensions depends on one dimensionless parameter, $\lambda_\phi/m_\textrm{pole}$. In the present work we have shown all results in terms of the respective pole mass. Hence, the above parameter simply relates to different couplings $\lambda_\phi$, measured in units of the pole mass $m_\textrm{pole}=1$. If interested in the massless limit of the theory, it is more convenient to keep the coupling fixed $\lambda_\phi=1$, and to depict all results for different pole masses. For example, in the work we have used  $\lambda_\phi=5,10,20$ with $m_\textrm{pole} = 1$, which can be read as $\lambda_\phi=1$ and $m_\textrm{pole} = 1/5, 1/10, 1/20$. This leads us to the spectral functions for propagator and four-vertex as well as the respective Euclidean correlators itself of the scalar field, depicted or rather redrawn in \Cref{fig:results_skeleton_c}. Rescaling the results in this way it gets clear the massless limit is readily investigated through the limit $\lambda_\phi \to \infty$. For a consistent treatment of the DSE, the vertices need to be well defined and have the appropriate scaling properties in this limit. This is given for the resummed $s$-channel four-point function that was introduced in \Cref{subsec:res4ptfct} and used in calculation of~\Cref{subsec:solution_coupled_system} and~\Cref{subsec:self_consistent_polarisation}.

Note that in the skeleton expansion as approximated in this work, the only other appearing vertex, which is the three-point function, is obtained directly from $\Gamma^{(4)}$ and hence has the same property. The resummation is hence suitable for studying the massless case in our skeleton expansion. In all diagrams except for the tadpole, all vertices are approximated at frequency zero. In consequence, they do not carry loop momentum and merely enter as multiplicative factor, which makes them well under control in the large coupling limit. For the tadpole however, this is not the case. Here, the four-point function enters via its spectral representation~\labelcref{eq:spec_rep_vert}. The additive classical contribution drops out by renormalisation, as it contributes momentum-independently. What is left is the vertex spectral function contributing to the tadpole loop integral, cf.~\labelcref{eq:tadpole_renormalized}. With increasing coupling, this spectral function gets larger, see~\Cref{fig:results_skeleton} or~\Cref{fig:results_skeleton_ex}. Since the multidimensional spectral integrals need to be evaluated numerically, a UV cutoff for the integrals needs to be chosen that minimises the error caused by doing so. This results in increasingly long-range integrals for the large coupling limit. The appropriate treatment of these integrals hence results in a technical obstacle, which will be the subject of a follow-up project. Apart from the technical aspect, this suggests that the tadpole diagram is the dominant contribution to the DSE and thus to the spectral function for $
\omega > 3 m_\mathrm{pole}$ in the $s$-channel approximation. As a result, one is left solely with polarisation and tadpole diagram, further simplifying the setup.

\section{Numerics} \label{app:Technical_details}

\begin{table}[b]
	\centering
	\begin{tabular}{|c ||   c| c|}
		\hline 
		\rule[1ex]{0pt}{1.5ex} \rule[-1ex]{0pt}{2.5ex}	 & \phantom{sp}Classical vertices \phantom{sp} &  \phantom{spaces12,} Skeleton \phantom{spaces12,}\\ 
		\hline
		\rule[1.5ex]{0pt}{1.5ex} 
		$g_{\text{pol}}$ &$-\frac{3}{2} \Gamma^{(2)}(\omega=0) \lambda_\phi$  & $-\frac{3}{2} \Gamma^{(2)}(\omega=0) \Gamma^{(4)}(\omega=0)$  \\[1ex]
		\hline & & \\[-1ex]
		$g_{\text{sunset}}$ & $- \frac{1}{6} \lambda_\phi^2$  & $\frac{1}{12} \Gamma^{(4)}(\omega=0)^2 $\\[1ex]
		\hline & & \\[-1ex]
		$g_{\text{squint}}$ & $\frac{3}{2} \lambda_\phi^2 \Gamma^{(2)}(\omega=0)$ & 0 \\[1ex]
		\hline & & \\[-1ex]
		$g_{\text{tad}}$ & 0  & $\frac{1}{2}$\\[1ex]
		\hline
	\end{tabular} 
	\caption{Prefactors of the propagator DSE diagrams in the different approximation schemes. The prefactors are obtained by the standard DSE prefactors and the loop-momentum and spectral parameter independent parts of the vertices. The tadpole factor in the approximation with classical vertices is set to zero, as the tadpole is absorbed completely in the mass renormalisation. The sunset prefactor in the skeleton expansion compensates the two-loop contributions of the tadpole with full four-vertex. In the self-consistent approximation of~\Cref{subsec:self_consistent_polarisation}, the skeleton prefactors apply with exception of the polarisation diagram, which is given by~\labelcref{eq:polGamma4}.}
	\label{tab:prefactors_DSE}
\end{table}
In order to compute the spectral integrals in \labelcref{eq:DSE_mink}, the integrands $I_j$ are discretised on suitable, evenly spaced momentum grids of usually around 100 points. The grids are chosen differently for each diagram such that peaked or discontinuous structures like the onset jumps in the imaginary parts are ideally resolved. The spectral integrations are performed numerically in \textsc{Mathematica} with standard global adaptive integration strategies using a relative precision goal of $10^{-3}$. All diagrams is interpolated separately in real and imaginary part in order to treat the sharp onset of the imaginary parts properly. All interpolations are performed using B-splines up to order 2 as all interpolants are real due to the split of real and imaginary part. The spectral function is then computed from the interpolated diagrams.

For a given set of parameters, convergence was usually reached within less than 10 iterations. \Cref{fig:specFuncConvPlot} demonstrates convergent behaviour. We estimate the relative precision of our routine to be  $\geq 10^{-4}$. Based on that, all mass poles shown in the right panel of~\Cref{fig:specFuncConvPlot}  left of the convergence boundary are indistinguishable. Normalising the scattering onsets of all spectral functions in the left panel to be identical, the continuous tail also converges pointwise beyond the convergence boundary, based on above precision estimate. 

\begin{widetext}

\section{Analytic expressions of the diagrams} \label{app:analytic_expressions}
Analytic expressions for the integrands in~\labelcref{eq:DSE_mink} before and after analytic continuation: \newline

\noindent \textbf{Polarization:}
\begin{equation}
\begin{split}
I_{\text{pol}}(p;\lambda_1,\lambda_2) \; = & \; \frac{1}{4 \pi  p}\arctan \left(\frac{p}{ \lambda_1 + \lambda_2 }\right) \,, \\[1ex] 
I_{\text{pol}}(\omega;\lambda_1,\lambda_2) \; = & \; \frac{1}{4 \pi \omega} \left[ \mathrm{artanh} \left(\frac{\omega}{\lambda_1+\lambda_2}\right)+ \imag \arg \left(1-\frac{\omega}{\lambda_1+\lambda_2}\right) \right]
\end{split}
\end{equation}

\noindent \textbf{Sunset:}
\begin{equation}
\begin{split}
I_{\text{sun}}(p;\lambda_1,\lambda_2,\lambda_3) \; &= \; \frac{1}{(4 \pi )^2} \Bigg( \frac{1}{2} \log \bigg(\frac{1}{(\lambda_1+\lambda_2+\lambda_3)^2+p^2}\bigg) - \frac{\lambda_1+\lambda_2+\lambda_3}{p} \arctan \bigg(\frac{p} {\lambda_1+\lambda_2+\lambda_3}\bigg) \Bigg) \,, \\[1ex]
I_{\text{sun}}(\omega;\lambda_1,\lambda_2,\lambda_3) \; &= \; \frac{1}{(4 \pi )^2} \Bigg( \frac{1}{2} \log \bigg(\frac{1}{(\lambda_1+\lambda_2+\lambda_3)^2-\omega^2}\bigg) \\ 
& - \frac{\lambda_1+\lambda_2+\lambda_3}{\omega} \bigg[ \mathrm{artanh} \bigg(\frac{\omega}{\lambda_1+\lambda_2+\lambda_3}\bigg)+ \imag \arg \bigg(1-\frac{\omega}{\lambda_1+\lambda_2+\lambda_3} \bigg) \bigg] \Bigg)
\end{split}
\end{equation}

\noindent \textbf{Squint:}
\begin{equation}
\begin{split}
I_{\text{squint}}(p;\lambda_1,\lambda_2,\lambda_3,\lambda_4) \; &= \; \frac{1}{(8 \pi)^2 \omega \lambda_4} \Bigg( 2 \log \bigg( \frac{\lambda_2 + \lambda_3 + \lambda_4}{\lambda_2 + \lambda_3 - \lambda_4} \bigg) \mathrm{artanh} \bigg( \frac{\omega}{\lambda_1 + \lambda_4} \bigg) + \imag \bigg[ \text{Li}_2 \bigg(\frac{ \imag p - \lambda_1 - \lambda_4}{\lambda_2+\lambda_3-\lambda_4}\bigg)  \\ 
&  - \; \text{Li}_2 \bigg( \frac{ \imag p - \lambda_1 + \lambda_4}{\lambda_2+\lambda_3+\lambda_4}\bigg) + \text{Li}_2\bigg( \frac{- \imag p - \lambda_1 + \lambda_4}{\lambda_2+\lambda_3+\lambda_4}\bigg) - \text{Li}_2 \bigg( \frac{- \imag p - \lambda_1 - \lambda_4}{\lambda_2+\lambda_3 - \lambda_4}\bigg) \bigg] \Bigg) \,, \\[1ex]
I_{\text{squint}}(\omega;\lambda_1,\lambda_2,\lambda_3,\lambda_4) \; &= \; \frac{1}{(8 \pi)^2 \omega \lambda_4} \Bigg( \text{Re} \bigg[ 2 \log \bigg( \frac{\lambda_2 + \lambda_3 + \lambda_4}{\lambda_2 + \lambda_3 - \lambda_4} \bigg) \mathrm{artanh} \bigg( \frac{\omega}{\lambda_1 + \lambda_4} \bigg) - \text{Li}_2 \bigg(\frac{ \omega - \lambda_1 - \lambda_4}{\lambda_2+\lambda_3-\lambda_4}\bigg)  \\ 
&  + \; \text{Li}_2 \bigg( \frac{- \omega- \lambda_1 - \lambda_4}{\lambda_2+\lambda_3-\lambda_4}\bigg) + \text{Li}_2 \bigg( \frac{ \omega - \lambda_1 + \lambda_4}{\lambda_2+\lambda_3+\lambda_4}\bigg) - \text{Li}_2\bigg( \frac{- \omega - \lambda_1 + \lambda_4}{\lambda_2+\lambda_3+\lambda_4}\bigg) \bigg] \\ 
& - \imag \; \theta \Big( \omega - | \lambda_1+\lambda_4 | \Big) \; \text{Im} \bigg[ 2 \log \bigg(\frac{\lambda_2+\lambda_3+\lambda_4}{\lambda_2+\lambda_3-\lambda_4}\bigg) \mathrm{artanh} \bigg( \frac{\omega}{\lambda_1+\lambda_4} \bigg) - \text{Li}_2 \bigg(\frac{ \omega - \lambda_1 - \lambda_4}{\lambda_2+\lambda_3-\lambda_4}\bigg) \\ 
& + \; \text{Li}_2 \bigg( \frac{- \omega- \lambda_1 - \lambda_4}{\lambda_2+\lambda_3-\lambda_4}\bigg) + \text{Li}_2 \bigg( \frac{ \omega - \lambda_1 + \lambda_4}{\lambda_2+\lambda_3+\lambda_4}\bigg) - \text{Li}_2\bigg( \frac{- \omega - \lambda_1 + \lambda_4}{\lambda_2+\lambda_3+\lambda_\phi}\bigg) \bigg] \Bigg)
\end{split}
\end{equation}
\end{widetext}

%

\end{document}